\documentclass{article}

\usepackage{float}
\usepackage{caption}
\usepackage{arxiv}

\usepackage[utf8]{inputenc} 
\usepackage[T1]{fontenc}    
\usepackage{hyperref}       
\usepackage{url}            
\usepackage{booktabs}       
\usepackage{amsfonts}       
\usepackage{nicefrac}       
\usepackage{microtype}      
\usepackage{lipsum}         
\usepackage{authblk}
\usepackage{graphicx}
\usepackage{natbib}
\usepackage{doi}
\usepackage{textcomp}        

\DeclareUnicodeCharacter{2212}{-}
\DeclareUnicodeCharacter{2265}{\ensuremath{\geq}}

\title{The Interplay Between Physical Activity, Protein Consumption, and Sleep Quality in Muscle Protein Synthesis}

\author[1]{\textbf{Ayush Devkota}\thanks{These authors contributed equally to this work.}}
\author[2]{\textbf{Manakamana Gautam}\thanks{These authors contributed equally to this work.}}
\author[3]{\textbf{Uttam Dhakal}}
\author[3]{\textbf{Suman Devkota}}
\author[5]{\textbf{Gaurav Kumar Gupta}}
\author[6]{\textbf{Ujjwal Nepal}}
\author[7]{\textbf{Amey Dinesh Dhuru}}
\author[4]{\textbf{Aniket Kumar Singh}\thanks{Corresponding author: \texttt{aniketsingh@ieee.org}}}

\affil[1]{Department of Applied Science and Chemical Engineering, Pulchowk Engineering College, Lalitpur, Nepal}
\affil[2]{School of Health and Allied Sciences, Pokhara University, Nepal}
\affil[3]{Department of Electrical and Computer Engineering, Youngstown State University, Ohio}
\affil[4]{Department of Computing and Information Systems, Youngstown State University, Ohio}
\affil[5]{Department of Computer Science and Information Systems, Youngstown State University, Ohio}
\affil[6]{Masters in Business Administration and Management, Maharishi International University, Fairfield, Iowa}
\affil[7]{State University of New York at Buffalo, Buffalo, New York}

\renewcommand{\shorttitle}{\textit{arXiv} Template}

\hypersetup{
pdftitle={The Interplay Between Physical Activity, Protein Consumption, and Sleep Quality in Muscle Protein Synthesis},
pdfsubject={q-bio.NC, q-bio.QM},
pdfauthor={Ayush Devkota, Manakamana Gautam, Uttam Dhakal, Suman Devkota, Gaurav Kumar Gupta, Ujjwal Nepal, Amey Dinesh Dhuru, Aniket Kumar Singh},
pdfkeywords={Myofibrillar protein synthesis, Muscle hypertrophy, Protein supplement, Resistance exercise, Quality sleep},
}

\begin{document}
\maketitle

\begin{abstract}
    This systematic review examines the synergistic and individual influences of resistance exercise, dietary protein supplementation, and sleep/recovery on muscle protein synthesis (MPS). Electronic databases such as Scopus, Google Scholar, and Web of Science were extensively used. Studies were selected based on relevance to the criteria and were ensured to be directly applicable to the objectives. Research indicates that a protein dose of 20 to 25 grams maximally stimulates MPS post-resistance training. It is observed that physically frail individuals aged 76 to 92 and middle-aged adults aged 62 to 74 have lower mixed muscle protein synthetic rates than individuals aged 20 to 32. High-whey protein and leucine-enriched supplements enhance MPS more efficiently than standard dairy products in older adults engaged in resistance programs. Similarly, protein intake before sleep boosts overnight MPS rates, which helps prevent muscle loss associated with sleep debt, exercise-induced damage, and muscle-wasting conditions like sarcopenia and cachexia. Resistance exercise is a functional intervention to achieve muscular adaptation and improve function. Future research should focus on variables such as fluctuating fitness levels, age groups, genetics, and lifestyle factors to generate more accurate and beneficial results.
\end{abstract}


\section{INTRODUCTION}

Protein synthesis and breakdown are dynamic processes that work together to regulate muscle mass, either building it up or breaking it down based on factors such as nutrition, exercise, and hormonal changes. Most mechanisms administering protein synthesis affect the initiation phase of mRNA translation [1]. If we eat more protein-rich foods, we will feed our muscles the required amino acids. These little guys then turn on and start building muscle [2]. Muscle protein synthesis is an important parameter of exercise adaptation and one of the most commonly used metrics for assessing the efficacy of various exercise and nutrition interventions [3].  
The rates of muscle protein synthesis (MPS) are accurately adjusted based on levels of physical activity, availability of nutrients, and overall health condition. In the absence of exercise, growth in MPS after feeding is momentary despite persistent precursor availability [4]. Resistance exercise keeps one fit and strong. It has been shown that this type of exercise can improve both physical fitness and metabolic health [5]. Resistance exercise induces a pronounced acute hormonal response. Such responses are more likely to be relevant for tissue growth and remodeling than chronic changes in baseline hormone concentrations [6]. Brain function is favorably influenced by moderate-intensity as well as high-intensity resistance exercise. Regular exercise can also enhance health status, functional capacity, and quality of life in older persons by reducing the risk of chronic diseases, as well as augmenting general well-being [7]. In patients with chronic heart failure, aerobic exercise is usually prescribed, while resistance exercises are oriented toward muscle force enhancement [8]. Resistance exercises have been indicated to be valuable in preserving muscle mass and bone mass during old age [9].

\begin{figure}
    \centering
    \includegraphics[width=0.6\textwidth]{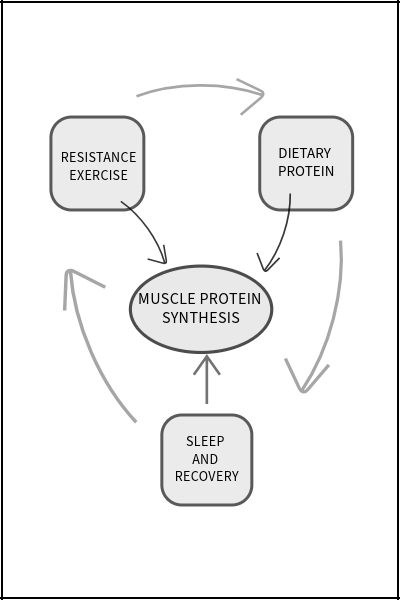}
    \caption{Inter-linkage between Resistance Exercise, Dietary Protein and Sleep with Muscle Protein Synthesis.}
    \label{fig:enter-label}
\end{figure}
 
It would be an excellent way of managing pain without straining the body too much. Not everyone can bear high-intensity aerobic exercises [10].
Very few studies have defined their conditions for resistance exercise in detail. They have usually focused on things like load magnitude, rest periods, and frequency of the sessions [11].
Research has consistently determined that resistance exercise can reduce anxiety in humans, with the most considerable and consistent reductions occurring after exercises of low to moderate intensity (less than 70\% of one repetition maximum) [12]. By 60, many individuals experience a significant reduction in muscle mass and strength, which can impact their overall health [2]. The decline in strength often seen with age is frequently attributed to decreasing muscle mass in older individuals [13]. Resistance training is a great way to boost muscle growth and improve overall function [14]. Older adults who are healthy may experience improvements in mitochondrial function and muscle strength after six months of resistance training to fix these issues (of weak muscles and less effective cells) in how they appear and fix them in their gene activity [15]. Even folks dealing with respiratory issues can see improvements in their muscle strength, physical abilities, and quality of life by following a good resistance training plan [16]. So, while just nutritional supplements might not completely stop muscle loss from cachexia, pairing them with an anabolic agent could help slow down or even prevent further muscle breakdown [17]. Also, it is super important to include various interventions like nutrition and exercise in plans for monitoring, evaluation, and discharge. These play a key role in preserving muscle mass and function in settings, be it clinical or community environments [18].

Sleep plays a big part in muscle recovery by influencing hormone secretion. Not getting enough sleep messes with blood hormones and cytokines that help with muscle recovery [19]. Sleep deprivation decreases protein synthesis pathways but develops degradation pathways, leading to muscle mass loss and making it tougher for muscles to recover from exercise, injuries, or conditions linked to muscle atrophy [20]. Plus, not getting adequate sleep is associated with a bunch of health issues like an increased risk of insulin resistance and type 2 diabetes [21]. Athletes need their beauty sleep since all the tough training can lead to sleep problems that mess with how much quality and quantity of sleep they get [22]. Sleep restriction is believed to be related to the loss of muscle mass in both humans and animals [23]. The belief that exercise enhances sleep is partially rooted in traditional theories suggesting that sleep functions to conserve energy, restore the body, and regulate temperature—all concepts that have heavily influenced research in this field [24]. Recent sleep research has broadened its focus from solely examining Central Nervous System (CNS) functions to include somatic physiology as well. The global regulation of sleep is not exclusively controlled by the CNS.  It is significantly affected by inputs from the entire body [25]. The primary approach directing skeletal muscle hypertrophy is a three-way mechanism: satellite cell activation, expansion or proliferation, and differentiation. Resistance exercise can create additional regulators that activate these cells and thus make contributory gains in muscle fiber size and number [26][27]. During childhood, skeletal muscles become larger primarily by hypertrophy - the expansion in size of individual muscle fibers. In contrast, muscle atrophy occurs from a state in which protein degradation rates become higher than protein synthesis rates [27].
Speaking about how vascular smooth muscle cells respond to growth signals, they can grow in size, increase their number, duplicate DNA without dividing, or even self-destruct [28]. In the initial postnatal phases, satellite cells in developing muscles proliferate significantly, contributing to a rise in myonuclear counts, as detailed by [29] in 2013. Also, skeletal muscles are pretty good at adapting to changes in the mechanical environment at both tissue and systemic levels. But, we are still figuring out exactly how mechanical signals turn into chemicals that affect muscle growth and metabolism [30]. 
Proteins are like the superheroes of cells, doing all sorts of important jobs like being tiny motors and helping with signaling. They also help out in chemical reactions, move stuff around, build up viral capsids, and create channels in membranes. They are super crucial for passing on genetic info from DNA to RNA [31]. Likewise, when amino acids don't get broken down in the small intestine, they take a ride through the portal vein to join in making proteins in muscles and other tissues [32].

Eating protein can rev up your energy use because your body has to work hard to make protein and urea and do gluconeogenesis. Diets with top-notch proteins that have all the vital amino acids can jack up energy use more than ones with lower-quality proteins [33]. The balance between making new proteins and breaking them down is key for growing protein levels in tissues because of continuous turnover [34]. Keeping your muscles healthy involves a back-and-forth between breaking down proteins (also known as protein turnover) when you're not eating (fasting) and building them back up when you chow down (feeding) [4]. Muscles can bulk up without turning on special cells (satellite cells) through changes in protein levels [29]. When comparing plant-based versus animal-based proteins, both seem pretty equal in how they affect calcium levels and bone health. Likewise, the deficiency in evidence that supports the superiority of plant-based ones over the other is limited in influencing these chemical mechanisms inside the human system [35].
People with different activity levels should aim for higher protein intakes. For those with minimal activity, it's recommended to have 1.0 g/kg BW/day. If you're moderately active, shoot for 1.3 g/kg BW/day. And if you're going intense, try to reach 1.6 g/kg BW/day.

It's generally safe for healthy adults to consume 2 g/kg BW/day of protein over the long term. However, don't go overboard – the upper limit is 3.5 g/kg BW/day for individuals used to high protein diets [32]. When it comes to building muscle and strength, athletes tend to gobble up more protein compared to their endurance-focused pals [36]. Leucine, in particular, is largely studied for its influential role in stimulating MPS. On the other hand, the impact of other EAA’s remains unknown due to limited data [37][36]. As people age, there is a noted decrease in how effectively MPS responds to essential amino acids (EAA’s), leading to anabolic resistance [37].
Studies show that after resistance exercise, the heart's parasympathetic modulation can take a hit in young, healthy individuals, leading to potential cardiovascular risks [38]. If you're diving into unfamiliar exercise territory with eccentric muscle contractions (lengthening), be prepared for muscle damage. You might notice decreased strength and range of motion, along with increased soreness and swelling. Keep an eye out for cellular proteins leaking as signs of damage [39].
After a grueling workout session, your immune system goes through various changes during recovery. Fancy this – interleukin-6 levels spike post-eccentric exercises, which signal muscle damage, which symbolizes multiple fascinating connections within the body [40].
Historically, inflammation was considered harmful to recovery from exercise, but it is now popularly identified that a well-regulated inflammatory response is crucial for effective muscle repair and regeneration [39]. Particularly, athletes who experienced intense exercise sessions leading to prolonged muscle fatigue were observed to have benefited from water immersion strategies [41]. For young youths, media exposure, peer pressure, and parental guidance all play a role in shaping muscle-building techniques [42].

In this review, we will analyze current research facts, detect gaps in the existing literature, and discuss practical recommendations for athletes and individuals who want to have a healthy lifestyle. Likewise, for individuals involved in physical training like resistance and endurance exercises as well. We will investigate the combined influence of resistance exercise, dietary protein supplementation, and quality sleep on muscle protein synthesis. Furthermore, we will examine their individual and synergistic roles in improving muscle growth, retention, and recovery. By combining these essential elements, this paper aims to offer detailed insights into optimizing muscle protein synthesis and improving overall muscular health.

\section{TRAINING PROTOCOLS OR METHODOLOGIES:
}
Muscle hypertrophy is the result of accumulated short-term increases in myofibrillar protein synthesis (MyoPS) after muscle damage has decreased [43]. In a human study, muscle protein synthesis rate (MPS) increases 50\% over baseline for 4 hours after heavy resistance training, and then, at 24 hours after the training session has finished, MPS is 109\% greater than at baseline, but MPS drops off so rapidly that by 36 hours it’s back to the 14\% baseline rate [44]. On the other hand, Dreyer et al. [45] found MPS to be 50\% for both men and women 2 hours after RT. Kumar et al. [46] found MPS to go back to baseline values in young (24 ± 6 years) and old (70 ± 5) men with the same BMI by 2-4 hours post-exercise.
In a study done back in the century, it was found that plyometric exercise increased mixed muscle protein FBR in the UT (untrained) group. Muscle protein FBR was unchanged in the T (trained) group. Exercise increased above resting in mixed muscle FSR in both groups. For Untrained: at rest by 0.036 ± 0.002 and exercised by 0.0802 ± 0.01. For Trained: at rest by 0.045 ± 0.004 and exercised by 0.067 ± 0.01(all datas in \%/h; P < 0.01). Besides that, exercise increased mixed muscle FBR by 37 ± 5\% (at rest by 0.076 ± 0.005 and exercised by 0.105 ± 0.01) in the UT group. It didn’t affect FBR much in the T group. Plyometric muscle contractions increase mixed muscle protein synthesis within 4 hours of exercise. Resistance training blunts this increase [47]. Aging is associated with marked declines in skeletal muscle protein content, muscle strength, muscle quality, and chemical modifications likely to impair protein function. Compared with young adults, frail 76 to 92-year-old men and women and healthy 62 to 74-year-old middle-aged men and women have lower rates of mixed muscle protein synthesis. The same amount of MHC and mixed muscle protein synthesis between frail, middle-aged, and young women and men increased after 2 weeks and 3 months of weightlifting exercise [48].

In the quadriceps, muscle protein synthesis fractional rate: Decline with age but is upregulated in older adults and young men and women after 2 weeks of resistance exercise. Before training, fractional muscle protein synthesis was suppressed in the elderly compared to the young but maximized at the same rate in both young and elderly after 2 weeks of exercise. When starting a resistance workout program, there's a spike in muscle protein production across all age groups without an increase in overall protein breakdown. If you're older, this uptick in muscle growth doesn't lead to more waste products in your urine – showing that your muscles aren't breaking down as much [49].
 In a Sprague-Dawley rats experiment, body mass was used to assign rats to RE, exercise control (EC), or sedentary cage control (CC) groups, and they were studied over 36 hours after 5 weeks of squat-like RE training. No differences in plantaris or soleus muscles’ acute measures of RPS (muscle protein synthesis) obtained 16 h after the last exercise session were registered at all (P > 0.05). Likewise, phosphorylation of 4E-BP1 bound it to eIF4E in all groups. Similarly, FSR was much higher in plantaris from RE compared to ECs and CCs when the cumulative response was established [50].
During three days, 12 healthy young men aged 22±1 were subjected to daily unilateral resistance-type exercise. Deuterated water dosing attained the level of steady body water enrichment of 0.70±0.03\%. There was an increase in the intramuscular free H-alanine enrichment before starting the exercise training to 1.84±0.06 MPE and it remained stable during the next three days of exercise training in both exercised and non-exercised control legs equaling 2.11±0.11 and 2.19±0.12 MPE, respectively. The above changes are not statistically significant (P>0.05). Muscle protein synthesis rates for the entire period averaged at 1.642±0.089\% per day compared to 1.984±0.118\% per day in the exercised leg indicating a distinct increase in the exercised leg (P<0.05) [51].
Hence, at the same relative intensity of exercise, an individual would be expected to exhibit hypertrophy and to alter the fed-state mixed muscle protein synthesis response to an acute bout of resistance exercise following 20 sessions of unilateral resistance training (eight weeks). One leg got trained and the other stayed untrained. In the type II fibers, the average muscle fiber cross-sectional area grew bigger after training in the trained leg by 20 ± 19\%. On the flipside in type I muscle fibers, it only increased in the trained leg by 16 ± 10\%. For acute resistance exercise, muscle protein FSR was up-regulated at 4 h in both legs. However, there was a more significant increase in the muscles of the trained leg compared to those of the untrained leg [52].
 
The study has shown that moderate resistance exercise may help to counteract the decline in MPS during bed rest. This exercise method also maintained muscle strength besides the prevention of reduced MPS. Bed rest is known to reduce muscle protein synthesis [53]. Low-volume, high-load resistance exercise does not induce more acute muscle anabolism than low-load, high-volume resistance exercise. They used three different loads/volumes: 90\% reps until you couldn't go on anymore (90 FAIL), 30\% matched in work to 90\% FAIL (30WM), or 30\% done until you couldn't carry on (30 FAIL). The synthesized MYO and MIX proteins rate significantly increased compared with rest at 4 hours following exercise but only at this time point for MIX at 24 hours when the workout was done at an intensity of 30 WM. At the same time, there was no change in these indices between exercises made under a workload of 30 WM and those carried out under a load of either 90 or 30 FAILS about MIX and MYO values after recovery from physical activity within four hours. It should be noted that type II fiber recruitment influences MYO protein synthesis after weight training independently on its level whereas training volume and muscle fiber activation rather than heavy loading per se determine this process [54].
Men and women show variations in their skeletal muscles. Men generally have larger muscle fibers than women. In older adults, muscle protein synthesis slows down compared to younger adults. It is unclear if this happens due to less activity or simply [45]. Young and elderly people underwent muscle protein synthesis studies before and following a 3-month exercise training program. What was surprising though, is that muscle protein synthesis did not increase post-training significantly for either group. The older group still had a 27\% slower rate. Interestingly, only the younger group had a 10\% rise in whole-body protein turnover. These results were unexpected since resistance exercise didn't boost muscle protein synthesis as anticipated [55].

 A single session of resistance exercise, whether pushing or pulling, can boost muscle protein balance for 48 hours. It doesn't matter whether the weights are lifted or brought down, but their effect does. The study checked participants at different times after the workout - when they were resting, and at three non-identical periods of 3 as well as 24, and 48 hours post-exercise. In comparison to rest, there was a substantial rise in MPS rate: 112\% at 3 hours, 65\% at 24 hours, and 34\% at 48 hours. The rate of breakdown was also influenced by exercise, with a jump observed at 3 hours (31\%) and 24 hours (18\%) before leveling back at the resting state by the 48th hour [56]. A detailed test was done to track myofibrillar protein synthesis following resistance training in the first, third, and tenth weeks of the program. Muscle damage peaked after the initial workout in week one but decreased gradually in weeks three and ten. Similarly, myofibrillar protein synthesis was highest in week one but lessened in weeks three and ten [43].
 When an individual hits the gym for some resistance training, the muscles grow bigger - that's muscle hypertrophy. Researchers studied this process for 16 weeks to see how muscle protein synthesis (MPS) activation connects with muscle growth from resistance training. MPS shot up by 235±38\% within 60–180 minutes after exercise and 184±28\% at about 180–360 minutes post-workout. The quads also got bigger by 7.9±1.6\% after all that sweat. Surprisingly, changes in quad size weren't linked to MPS rates within 1–6 hours post-workout as per an experimental analysis. How much MPS happens right after working out doesn't predict long-term muscle growth according to these outcomes [57]. This aligns with the findings of another study [58], which showed no clear connection between muscle protein synthesis and muscle growth following the first weight training session. In the study of [57], participants chugged a protein drink after each workout, while in the study of [58], they skipped the post-workout snack altogether.

\begin{figure}
    \centering
    \includegraphics[width=0.9\textwidth]{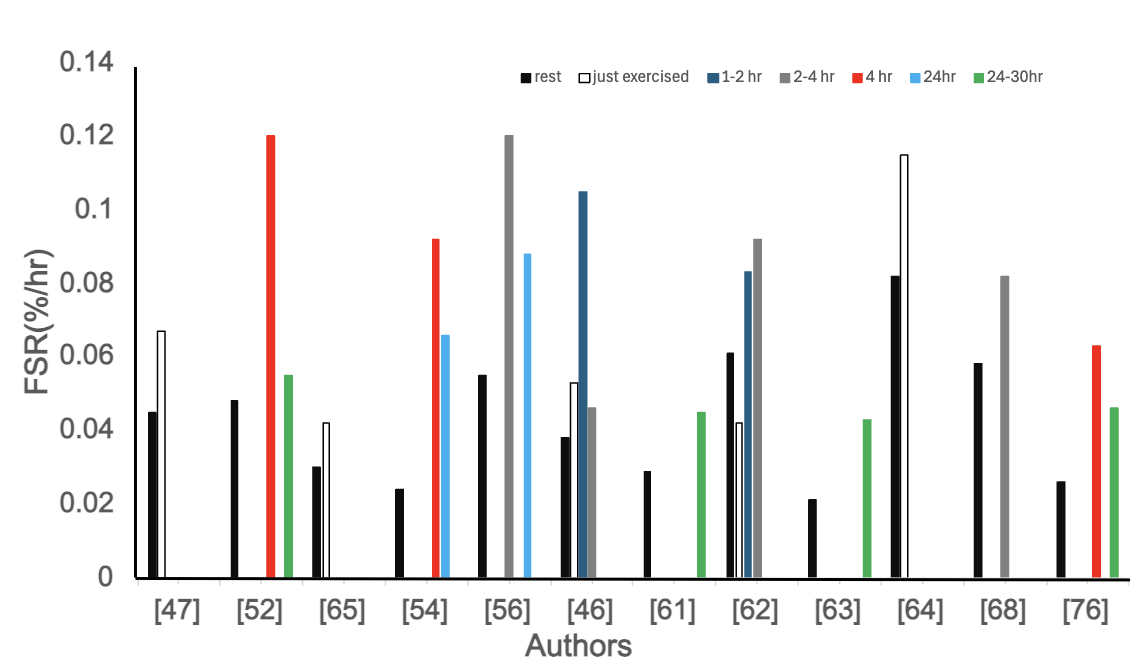}
    \caption{This schematic bar diagram represents the rate of Fractional Synthesis at different time frames, as demonstrated by multiple authors. Data from [47] are for plyometric exercises in trained individuals. [52], [62], [65], and [64] are for the trained leg. [54] is for 30 FAIL conditions. [56] is for Concentric and Eccentric exercises. The data from [46] are evaluated from young individuals. Likewise, [61] are from 3 working sets. [63] data are from SLOW(6 secs) repetitions. Data from [68] are for blood flow restriction. [76] are for 5 minutes rest time. 
}
    \label{fig:enter-label}
\end{figure}

Experimentation to identify how doing resistance-type exercises when a person is in a fed state boosts the rate at which your muscles make protein has some fascinating insights. A group of 10 guys who weren't regular gym-goers did some leg exercises individually. Those who worked out had a magnificent 20\% higher rates of MPS compared to those who didn't. Surprisingly, the AMPK phosphorylation (a term for a chemical process) wasn't affected by the exercise when they had just eaten. However, another kind of phosphorylation called 4E-BP1 was lower by almost 20\% in the guys who worked out [59]. Moving on, when older folks in their 70s and 80s do resistance exercises, there's an exponential augment in muscle protein synthesis rates. A study looked at different protein synthesis rates in young adults of age 23-32 and seniors of age 78-84 before and after a 2-week lifting program. After the program, the younger group saw an 88-121\% increase in their muscle protein synthesis rates, while the older group saw a boost of 105-182\%. As we age, certain muscle protein synthesis rates decrease more than others. But here's the cool part - even seniors can improve their muscle protein synthesis rates with short-term resistance exercises, just like younger adults can. Whereas, measurement variability of actin protein synthesis rates is way more substantial than measuring MHC or mixed muscle proteins [60]. 
 An experiment to assess how muscles grow after exercise in males and females found that muscle protein synthesis (MPS) jumped up by 47\% in women and 52\% in men after some intense leg workouts. This all happened during the first two hours after the exercise. What's interesting is that this boost in MPS doesn't seem to care if you're a guy or a gal. It doesn't affect how leg muscles grow in young folks at all.
When they looked at how MPS worked when folks were resting state, no big differences appeared between young guys and gals [45]. Now, when it comes to older men hitting the training, things change a bit. Their muscles don't respond as well to exercise signals. Instead of a smooth increase with more intense exercise, as the young ones show, it's more like a slow start at 20-40\% effort( of their 1 repetition maximums), then a big burst at 60\% effort without much happening past 90\%. The response from older guys is smaller than what younger folks see.
To add to that, MPS drops big time for both age groups - whether they're 24 or 70 years old - who have similar BMI numbers. After just a few hours post-workout, their muscle protein synthesis drops back close to regular levels [46].

An experiment was done to see any differences between doing just one set of resistance exercises and doing three sets (3SET). The results showed that after doing one set, muscle protein synthesis (MPS) went up at 5 hours and then returned to normal levels at 29 hours. But when it came to doing three sets, the increase in MPS was even better at 5 hours and 29 hours after the workout. This means that doing three sets of resistance exercises can help you build more muscle over time compared to just doing one set.
Interestingly, feeding didn't seem to make a big difference in MPS after doing only one set of exercises between 24-29 hours later. It seems like you need to do a certain amount of exercise to get your muscles ready for food [61]. Other studies have also shown that resistance exercises are great for boosting MPS and muscle growth. Your body starts making more protein within a few hours of working out and this increase can last up to two days.
As the study was done with 11 participants (seven men and four women), researchers found that muscle protein synthesis increased significantly at 1 and 2 hours after exercise but dropped during the workout itself. These findings suggest that pushing yourself with multiple sets of resistance exercises may be the way to go if you want to see some serious MPS and muscle gains [62].
Lifting weights boosts how muscles grow, no matter how long you lift. The impact is much bigger when holding the muscles for longer periods at first. Doing leg extensions at 30\% of max speed - slowly going up for 6 seconds and down for 6 seconds until tired - leads to more muscle growth compared to doing it fast (1 second up and 1 second down). This shows that spending enough time straining your muscles during workouts is key to building muscle. Keeping the muscles under tension during weightlifting didn't boost immediate muscle protein synthesis, but it did lead to a delayed growth spurt after 24-30 hours of rest with extra protein intake. Plus, there were consequential augments in mitochondrial protein production during the recovery period of 24-30 hours in both slow and controlled conditions [63].

When an individual performs exercises like endurance (EE) and resistance (RE), they can help stimulate mixed skeletal MPS. RE \& EE trigger some cool stuff in your muscles -- they make myofibrillar protein grow and expand your mitochondria. This means your muscles get stronger and more efficient at producing energy. In people who are new to training, RE can boost myofibrillar and mitochondrial protein synthesis by around 67\% and 69\%, respectively (that difference is pretty big!). After training, only myofibrillar MPS gets a boost from RE, about 36\%. On the other hand, EE boosts mitochondrial protein synthesis whether you're trained or not (by 154\% and 105\%, respectively). But interestingly, it doesn't do much for myofibrillar MPS after 10 weeks of training. This shows that single-leg endurance exercises don't help with myofibrillar protein synthesis, even after training [64].
A different study [65] found that resistance training changes how your muscles respond to exercise by increasing myofibrillar protein production while decreasing non-myofibrillar proteins. They looked at people doing eight weeks of unilateral resistance exercises and saw their muscle fibers get bigger. Even though resting mixed MPS increased by 48\%, myofibrillar MPS stayed the same between trained and untrained folks, eventually going up in both groups. When they [52] checked out the effects of resistance training on quadriceps muscle, they noticed that the protein synthesis rate was higher in the trained leg compared to the untrained leg.
Another experiment [66] had people doing heavy resistance exercises on their biceps while one arm worked out and the other provided support or acted as a control arm. They wanted to see how much muscle protein was influenced by some natural biological processes like transcription and translation. Turns out, MPS went up a lot more in the worked-out biceps compared to the chill arm in both groups (they were like 0.067\% vs 0.1007\% per hour; and 0.0452\% vs 0.0944\% per hour). RNA activity also shot up in the worked-out biceps but stayed steady in the non-worked biceps for both groups.

Resistance exercise training is super beneficial for frail women and men over 76 years old. 8 women and 4 men, who were slightly weak physically, joined a 3-month weight-lifting program after an initial 3 months of physical therapy. In the supervised resistance exercise group, the production of muscle torque in the knee extensor was boosted. 
For elderly folks aged between 76 to 92 with delicately balanced physical strength, their muscle synthetic pathways react positively to increased contractile activity from progressive resistance training. It's like your muscles a little push to be stronger and healthier [67].
Adding blood flow restriction (REFR) to low-intensity resistance exercise can help increase muscle strength and size just like routine methods with heavy loads do. Study participants were analyzed both with and without blood flow restriction during exercises. The muscle fractional synthetic rate went up by a whopping 46\% after REFR exercises compared to control exercises without blood flow restriction [68].
In another study, twenty-three sedentary men around 67 years old engaged in a 16-week progressive resistance training regimen. They wanted to see if injecting growth hormone (GH) along with exercise could boost muscle protein anabolism. Interestingly, in the GH group, aspects like fat-free mass, total body water, protein breakdown, and synthesis rates were significantly increased. Likewise, growth in vastus lateralis MPS rate, training-specific isokinetic and isotonic muscle strength, and urinary creatinine excretion were pretty much similar in individuals.
While resistance exercise did wonders for muscle strength in older men, combining it with GH didn't show a greater impact on muscle anabolism or strength. It seems that the escalation in noncontractile protein and fluid retention might have contributed more to boosting fat-free mass with GH treatment than direct effects on muscles themselves [69]. 

When an individual performs resistance exercise until they can't go on anymore, it boosts their muscle response for a whole day. Eating protein helps speed up muscle growth rates ( accountable for influencing rates of myofibrillar MPS rates over rates of fasting state by 0.016 ± 0.002\%/h) after exercise and keeps the response surged for 24 hours, especially in specific conditions. In a study, 15 guys checked out muscle growth after resistance training for 24 hours or after protein intake while resting. They did leg exercises at different intensities - 30\% or 90\% of their max strength till failure [70].
As we age, losing muscle mass and strength shoots up the risk of falls and reliance on others. Research points out that older adults have trouble synthesizing muscle protein after intense exercise, but not much is known about its breakdown in them. Muscle samples were taken from both younger (around 27 years old) and older (about 70 years old) folks before and up to 24 hours after exercising intensely. Some proteins went up while others went down in both groups post-exercise, indicating no major differences in muscle breakdown between them over time. MuRF1 mRNA expression was enhanced, whereas GABARAP mRNA declined after RE in the trial of younger and older adults. Old age doesn't seem to affect how muscles respond to exercise stress, so enhancing protein synthesis could be key in preventing muscle loss or avoiding harmful sarcopenia [71].
For seniors looking to stay fit and have physical strength, resistance, and aerobic exercises are recommended. High-intensity interval training (HIIT) might also be a good replacement for aerobics for them. A study involving inactive older men showed that both resistance training and HIIT increased muscle protein synthesis rates for up to two days after the workout compared to just resting. Interestingly, HIIT boosted one type of protein synthesis more than resistance training did - showing promise as an effective exercise routine for aging men. HIIT was the only mode of exercise to augment sarcoplasmic protein fractional synthetic rate 24 hours post-exercise ($2.30 \pm 0.34\% \, d^{-1}$ vs. $1.83 \pm 0.21\% \, d^{-1}$). This explains why HIIT greatly influences rates of myofibrillar along with sarcoplasmic fractional synthesis in the age group of 67 ± 4 years [72].

In a trial of 9 lean and 8 obese active young adults, it was found that lean folks show a stronger response to nutrient stimulation, especially after resistance exercise.
Obese individuals tend to have higher baseline levels of markers like plasma triacylglycerol, serum insulin, plasma CRP, and plasma cholesterol. This can lead to increased insulin resistance. On top of that momentary advances in the intracellular signaling protein's phosphorylation like p70S6K, AKT, and 4EBP1 inside the trained leg were observed. However, when it comes to muscle protein synthesis rates, both lean and obese participants didn't show significant differences at rest or after exercise [73].
During aging, maintaining skeletal muscle is crucial for avoiding fractures and falls. A study found that old muscles may not activate certain signaling pathways as effectively as younger muscles after resistance exercise. This is delayed due to old muscles being unable to activate both ERK and mTOR after resistance training. MPS was enhanced later in old (3-6 hours) and earlier in young (1-3 hours) participants. At 1 hour post-exercise, ERK1/2 and MNK1 phosphorylation augmented, whereas eIF2 phosphorylation reduced in the younger ones only. On the other hand, MNK1 phosphorylation was lower at 3 hours. Likewise, AMPK phosphorylation was greater in the elderly 1-3 hours post-exercise  Muscle protein synthesis improved later in old participants and earlier in young ones [74].

In a 16-week trial, 23 adults did RT protocols. After 15, 30, and 60 minutes following the first and last training sessions, serum cytokines and hormones were measured right away after the study.
Changes in IL-6 due to exercise are linked with hypertrophy. However, not much is known about how IL-6 plays a role in hypertrophy. The variations in muscle AR (androgen receptor) protein content \& spikes in p70S6K phosphorylation are tied to muscle hypertrophy, indicating the inner muscle part of overall processes for helping with hypertrophy [75].
A study a few years ago found that with one minute of rest(short) between sets during resistance training, myofibrillar protein synthesis was lower in the early recovery compared to 5 minutes of rest. Furthermore, myofibrillar protein synthesis increased by 76\% and 152\% at 0-4 hours post-exercise in the 1-minute and 5-minute inter-set rest periods, with noticeably better results in the group that rested for 5 minutes between sets. Myofibrillar protein synthesis rates were elevated above baseline levels for 24-48 hours after exercise [76].

\section{PROTEIN SUPPLEMENTATION:
}
According to various studies [77], [78], [79], and [80], about 20-25 g of protein is needed to get the best boost in muscle protein synthesis (MPS) in our muscles after a workout. A study by [78] found that when you have more than 20 g of whey, ureagenesis and amino acid oxidation get going. They tested 0, 10, 20, and 40 grams of whey protein after an intense training program.
In another study [79], they wanted to see how our bodies react to different amounts of protein (MPS) and albumin protein synthesis (APS) after working out. They tried drinks with 0, 5, 10, 20, or 40 g of protein(from eggs). They checked protein synthesis and leucine oxidation for four hours post-workout. MPS was highest at 20 grams, and leucine oxidation went up significantly after having 20 or 40 grams of protein.
Furthermore, [80] advised getting at least 0.4 g/kg/meal for protein over four meals each day to reach a total of 1.6 g/kg for the most gains in a day. They also agreed with [78] that going beyond 20-25 grams of high-quality protein leads to making urea or getting turned into energy.
In a study from the past decade [81], taking in either 15 or 30 g of protein after a resistance program during an energy deficit boosted MPS by either a 16\% or 34\% above normal energy balance. Combining resistance exercise with more protein helps increase muscle building during short-term energy deficits and might help maintain muscle mass in the long run after a good workout session. During short-term energy deficits, muscle building goes down but doing resistance exercise during this time brings it back up to normal levels like when you're well-fed.

In previous studies, protein quantity is essential for MPS. Specific intake pattern's effects on MPS over 12 hours are less known. A study on 24 trained men who had 80 grams of whey using 3 protocols was conducted. They consumed 10 grams 8 times every 1.5 hours (PULSE) 20 grams 4 times every 3 hours (intermediate: INT), or 40 grams 2 times every 6 hours (BOLUS). It was found that consuming 20 g of whey protein every 3 hours (INT) was best for stimulating MPS in the 12-hour window compared to the other two methods. This study sheds light on how varying protein intake distribution affects muscle anabolic responses, potentially improving resistance training outcomes for muscle growth [82].
Researchers discovered that rapidly increasing amino acids after exercise boosts MPS and anabolic signaling more than slowly digesting protein consumed in small pulses. A single BOLUS (25 g dose) increased MPS by 95\% at 1-3 hours and 3-5 hours by 193\% compared to PULSE (10 drinks of 2.5 g each every 20 minutes), leading to rates of 42\% and 121\%, respectively. Resistance exercise aims to maintain high myofibrillar protein consumption rates purposely [83].
The authors were on a quest to find whether consuming protein two hours post-training session (P2) made a difference compared to immediate consumption post-exercise (P0). Surprisingly, no significant hypertrophy difference was noticed in the P2 group after the resistance exercise program for 12 weeks compared to the P0 group. This highlights the importance of early protein intake timing for muscle recovery and net synthesis. For m. quadriceps femoris, CSA (cross-sectional area) along with the mean area of fiber enhanced in the P0 group. $54.6 \pm 0.5$ to $58.3 \pm 0.5$ cm\textsuperscript{2} and $4047 \pm 320$ to $5019 \pm 615$ $\mu$m\textsuperscript{2}, respectively, but the P2 group observed no influential changes [84].

Recent findings indicate that consuming zero grams of Protein resulted in MyoPS rates being stimulated by only half (46\% less) as much as with 30 grams of Protein. For maximizing MyoPS rates after endurance exercise, just a mere amount of 30g of Protein is sufficient. However, protein intake did not increase mitochondrial protein synthesis rates. To boost MyoPS rates during convalescence from a single endurance exercise session, consuming only 30g of protein is reported to be adequate [85].
\begin{figure}
     \centering
     \includegraphics[width=0.8\textwidth]{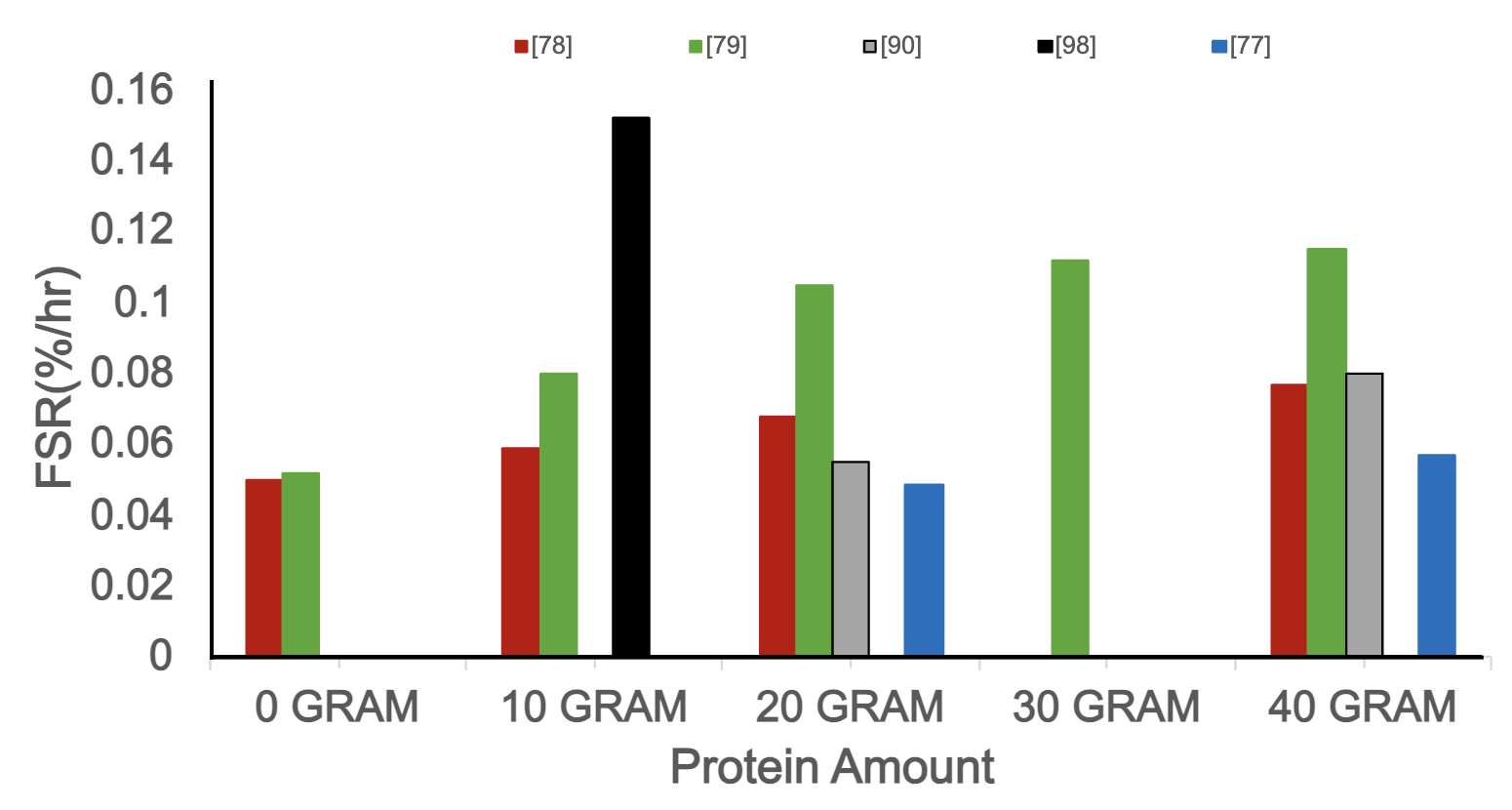}
     \caption{This schematic bar diagram represents the rate of Fractional Synthesis at multiple Protein Quantity. The data from [78], [79], [90], [98], and [77] are from exercised conditions.}
     \label{fig:enter-label}
 \end{figure}
 
In an examination done with 24 folks (both guys \& girls) by [86], MPS didn't change right away after a tough resistance workout with caseinate and whey consumption in older individuals during the first 6-hour recovery time. What the authors found was that there wasn't any difference between snacking on caseinate 30 minutes before or right after exercise when it came to myofibrillar protein synthesis rate, and the same goes for MPS compared to using caseinate before or after training. They weren't sure if gobbling up protein before or after lifting weights made a big difference in MPS. Their findings lined up with [87], who noticed a muscle-building response to munching on 20 g of whey proteins before or an hour after pumping iron (weight training), and no distinctiveness was identified between time points.

A wee bit of 6.25 g of whey protein mixed with essential amino acids or just leucine did just as well as full-on protein (WHEY) in getting MPS going after working out, but only WHEY (25 g) over 3–5 hours post-exercise (184\%, LEU 55\%, and EAA-LEU 35\%) was enough to keep that MPS rolling high after exercising. The results here match up nicely with WHEY's MPS over 3-5 hours observed in [83]. The MPS for both LEU and EAA-LEU had fallen back down to regular levels, so WHEY seems like the best bet for boosting post-workout muscle protein accretion [88]. Having a nice bit (25–30 g) of top-notch protein at every meal along with getting some exercise around mealtime could be one solid way to keep those muscles strong and healthy for older folks throughout their bed rest. Plus, taking action carefully against fast muscle loss during tough times and less physical activity by bringing in some nutritional help like amino acids, and extra protein, and mixing in some physical therapy moves is a smart move/ a great tactic [89].
 Whey protein seems to be superior at boosting muscle protein synthesis. Not only is better than soy, but it also beats out other sources of dairy protein like intact or hydrolyzed casein. Muscle protein synthesis gets a boost from having more amino acids available. And get this - younger muscles respond better to amino acids' muscle-building effects compared to older muscles.
In a study with 30 men, W20 showed higher rates of synthesis of muscle protein than S20. The rates were about the same in both exercised and unexercised leg muscles, and they were not varying from 0 g (P = 0.41. The study also found similar results for S40 and W40 [90].

In another study back in the mid-2000s, participants took either 20g of protein (6g amino acids, 14g casein, and whey) or a dextrose placebo before and after exercising to meet the daily 40g protein requirement. The folks who had the protein supplement saw improvements in muscle strength, fat-free mass, body mass, serum IGF-1, IIa expression, and myofibrillar protein.
They recommended resistance training for a 10-week period while having 20 grams of protein along with amino acids an hour before and after working out. This combo worked better than a carbohydrate placebo at boosting markers of MPS and muscle performance improvements [91].
As we age, our muscles can weaken and shrink unless we build up our proteins through resistance exercises(incorporating a net synthesis of muscular proteins). While post-exercise amino acid supplements play a role in muscle synthesis, the timing of your protein intake is still up for debate[84].
Another study [92] found that taking 25g of whey protein after resistance exercise boosted Myo MPS in rested legs after 3 hours by a whopping 163\%! In contrast, exercise legs showed increased MYO protein synthesis at 1, 3, and 5 hours compared to being fast. MYO protein synthesis in the exercised leg was stimulated above FAST by 100, 216, and finally by 229\% at 1, 3, and 5 h, with the escalation at 5 hours being superior to FED, which recommends sarcoplasmic and myofibrillar protein synthesis is temporarily stimulated by protein uptake.
Hence, taking in more protein helps offset muscle loss when an individual is not active enough and speeds up recovery when they get active again [93].
In a study that lasted for 14 weeks, researchers examined how resistance training (RT) paired with timed consumption of isoenergetic protein versus carbohydrate supplementation impacts muscle growth and strength. The results showed similar enhancements in strength for both groups. However, only those who received protein saw an increase in muscle size (18\% ± for type I and 26\% ± 5\% for type II), while the carbohydrate group did not show significant improvements beyond their initial levels. This suggests that just taking carbohydrates without combining them with resistance training may not lead to long-term muscle development or anabolism [94].

Interestingly, non-essential amino acids (NEAA) are not essential to boost net muscle protein balance. Instead, there is a direct relationship between essential amino acid (EAA) intake and muscle protein production. Consuming a modest amount of EAA (6 g) can effectively stimulate net muscle protein balance following resistance exercise. Surprisingly, ingesting the drink containing EAAs two hours after training had a similar impact on muscle protein synthesis as the drink consumed one hour post-workout. These findings indicate that increasing arterial and interstitial amino acid concentrations is more crucial than the absolute amino acid levels for stimulating MPS [95].
In a detailed observation, eight men underwent resistance training while consuming either a carbohydrate beverage (CHO: 10 g maltodextrin, 21 g fructose) or a whey protein drink (WHEY: 10 g protein, 21 g fructose). The group that drank whey protein along with carbohydrates showed an increase in muscle protein synthesis after their workout session compared to the carbohydrate-only group. Athletes commonly use whey protein as a supplement to help build muscles, although there is limited evidence supporting its effectiveness in stimulating desired muscle growth [96].
Furthermore, research from the late 2000s highlighted the positive effects of combining protein with carbohydrates to enhance whole-body net protein balance after intense aerobic activities (1.2 g CHO, 1.2 g CHO + 0.4 g protein, or 1.6 g CHO; unit in kg$^{-1} \cdot$h$^{-1}$) in random order. The trial was performed with 6 healthy men consuming drinks during the initial 3 hours of recovery. During recovery after exercise, consuming protein along with carbohydrates increases muscle fractional synthesis rate and improves overall net protein balance within the body. Conversely, the impact of altering nutrient intake on muscle protein dynamics following aerobic exercise remains largely untapped. While the exact proteins responsible for this boost in muscle synthesis were not identified, this study was pioneering in directly examining how manipulating carbohydrate and protein intake post-endurance exercise could impact muscle synthesis rates using needle biopsy techniques [97].

In a scientific study, 3 groups of 6 each did leg resistance exercises. They also consumed a beverage with the same profile of essential AA (10 g) as whey, micellar casein, or soy protein. The mixed MPS at rest was higher with faster proteins and there was a significant difference between whey and soy as well as whey and casein, but there was no significant difference between whey and a mixture of soy and casein. Whey had a higher value of MPS as compared to casein which had 93\% more MPS and soy, which had 18\% more. After exercise (whey > soy > casein), whey enhanced MPS by 122\%  than casein and 31\% than soy [98].
Another study with 22 young squads explored if ingesting amino acids \& carbs before exercise boosts muscle protein synthesis. One group (each with 11) fasted, while the other downed a solution of EAAs and CHOs an hour before starting the training session. Muscle FSR rose right after ingestion, dipped during exercise but remained unchanged for 1 hour post-exercise for the EAA CHO group. FSR dropped during exercise for the fasting group but rose post-exercise too. However, the 2-hour post-exercise FSR was enhanced by 50\% in both groups, with no variation among groups. It supports the hypothesis that EAA CHO consumption following weight training is more effective at enhancing muscle protein synthesis throughout post-exercise recovery than before resistance exercise, which does not enhance recovery period muscle protein synthesis [99].

So, they did this study to see how different amounts of protein in our diet affect our body composition and muscle protein synthesis when we're trying to cut back on calories. In an experiment with 39 adults eating different protein levels during energy deficit for 31 days. During energy deficiency, the protein diet had an anabolic response similar to WM (Weight maintenance) and 2X-RDA/3X-RDA(recommended dietary allowance) but lower RDA levels. The author reported that consuming dietary protein at levels surpassing the Recommended Dietary Allowance (RDA) may help preserve fat-free mass throughout short-term weight loss [100].
High protein intake didn't stop lean mass loss in college students(allocated into an energy-restricted group; 30 kcal/kg or a eucaloric control group; 45 kcal/kg) on energy restriction for six weeks. Uncertain if a protein intake of 2.8 g/kg prevented more muscle loss during weight maintenance. This form of energy restriction doesn't mess with your muscle contractility. The underlying cause of these benefits attributed to the high protein intake remains unclear and requires further investigation. Both groups had less energy, but there were no changes in their mood [101].

\begin{figure}
     \centering
     \includegraphics[width=0.9\textwidth]{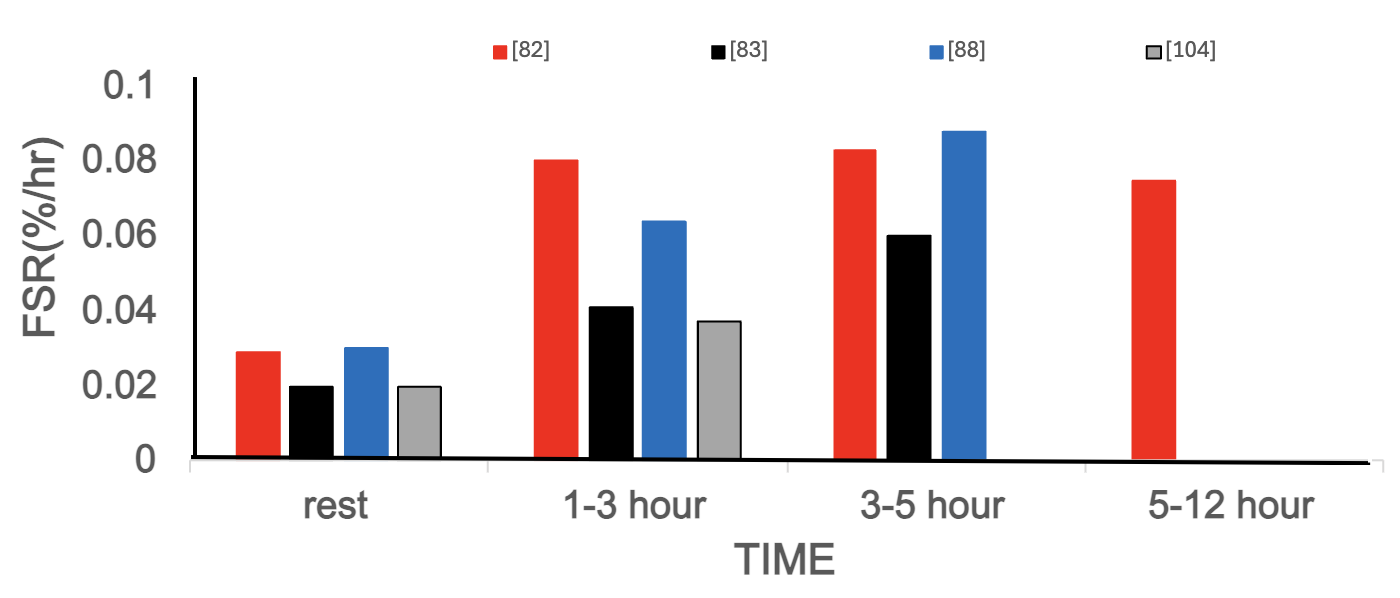}
     \caption{This schematic bar diagram represents the rate of Fractional Synthesis at different periods from 20-25 gram high-quality protein. }
     \label{fig:enter-label}
 \end{figure}

In an extensive analysis, 8 aged individuals (75 ± 1 years) and 8 young ones (20 ± 1 years), all slender, were split into two teams. Each team either chowed down on just carbs or carbs plus protein along with leucine after doing their regular tasks for 30 minutes. Samples over 6 hours revealed that both squads were losing protein (protein balance was negative) when they underwent carbohydrate consumption. Contrary to this, when they added protein and leucine to the mix, the protein balance turned positive. 
So, the rates of making muscle proteins were way less( Mixed MPS) in the carb-only deal compared to the carb-protein-leucine combo for both the young guns and wise old companions. However, the surge in muscle protein making was the same for lean young men as well as older ones [102].
Sarcopenia, which is like weakening muscles with age, might be linked to not making enough muscle protein after eating or working out. This weak response could be because proteins aren't getting digested and absorbed properly, leading to fewer amino acids floating around after meals. But here's a twist - digestion, and absorption of dietary protein don't tank(aren't compromised) after exercise or as we get older.
Also, engaging in exercise before chowing down on protein facilitates your body to use those amino acids better to make new muscles in young and old men. A rapid rise in amino acid levels was seen in both groups after eating protein, with heightened levels found in the oldies than in the young ones [103]
As reported by a few intellectuals, sixteen average middle-aged youths ate either milk protein (8 men) or whey protein (8 other men) while having 13C6 phenylalanine pumped into their veins. Turns out, muscle protein FSR went back to normal within a few hours (90 to 200 minutes) for both groups after eating protein.

Even though whey protein helps with digestion and leucine availability more than milk protein does, both(20 g) types led to similar muscle-making boosts in middle-aged men [104]. In a sweet clinical test run, people first balanced their energy intake for a week, then reduced it for another week (500 kcal per day energy restriction), followed by even less movement(<750 steps per day) for two weeks before going back to normal routine for a week.
They gulped down 1.6 grams of protein( in 24 hours) per kilogram of their respective body weight daily from food and supplements( around 45\% ± 9\% ) like whey protein or collagen peptides twice a day (30 grams). Muscle forming slowed during reduced energy intake but bounced back during recovery only for the whey protein group.
Adding supplements didn't guard against losing leg lean mass except for whey protein helping out during recovery from being inactive energy state only adds benefit [93].
So, in a 12-week trial, ECC and CONC resistance training beefed up the muscles with the whey protein and carbohydrate group compared to just the carbohydrate group. The muscle growth was way more (7.3 ±1.0 versus 3.4 ± 0.8\%), no matter what exercise type was done. The author suggests that doing those eccentric contractions (ECC) can help muscles grow more/ stronger driver of muscle growth, than concentric contractions (CONC), especially when combined with some protein-boosting [105].

Having some protein along with carbs throughout the exercise and after exercise can augment protein synthesis in the entire body. But here's the funny thing, it doesn't boost MPS rates during the overnight recovery time of 9 hours. Overall, throughout exercise, both whole-body and MPS rates jump up by 29\% and 48\%, respectively, when protein and carbs are in the mix. The synthetic rates during exercise were about 0.056 ± 0.003\%/hour in one group and 0.083 ± 0.011\%/hour in another group. Then, after that workout sesh, whole-body protein synthesis was bigger by 19\% in one group compared to the other. Also, the average MPS rates during the 9-hour overnight recovery didn't show any significant variations between the groups. It was 0.056 ± 0.004\% per hour for the C+P group, and 0.057 ± 0.004\% per hour for the W group. [106].
In a detailed study with 29 females and 37 males, they looked into how a year of protein boosting affected muscle synthesis in healthy older Danes from Denmark - It turns out, that nothing has changed in a year on some aspects of muscle's protein synthesis along with its metabolome. After a good while of letting a new balance of body chemistry set in, it seems that eating more protein did not change the basic and after-meal muscle growth rate [107].

They tried another experiment too to check out how much protein could help with muscle growth. A smaller serving with less protein(30g in 113 g serving) revved up MPS by half but going bigger(90g in 340g serving) didn't do much more for both young and older folks [108]. In terms of stimulating muscle growth, it seems likely that under resting or non-exercising conditions, consumption of more than 30g of protein at once isn't needed for muscles to grow big and strong. 
In a study from [109], it was found that adding protein during weight loss helped maintain thigh muscle volume after 5\% weight reduction and lowered after 10\% weight deficiency in middle-aged women suffering from obesity. It's suggested to eat high-protein, leucine foods to protect muscle mass when losing weight, but how well this works isn't clear yet. Another study [110] gave older adults a protein supplement enriched with leucine or an iso-caloric milk drink with 6 g protein after exercise, and it found it increased muscle protein synthesis compared to a regular dairy product.
Another trial on rats showed that whey-based protein supplements increased protein synthesis more than carbohydrates after exercise. The supplements comprising whey-based proteins (Amino Acids and Whey) led to a significant rise in the fractional rate of protein synthesis (FSR) than the one with CHO. This suggests that certain peptides in whey protein are better than amino acids for building muscle [111].

\section{THE ROLE OF SLEEP AND RECOVERY
}
Sleep is super important for muscle recovery because it affects hormone secretion. When you don't get enough sleep or have limited sleep, it messes with your blood hormone levels and cytokines, which can mess up your skeletal muscle recovery [19].
In a recent journal article, they found that taking a protein (PRO) supplement before bed with 40g of casein protein didn't make a big difference in how well your muscles the next day or how hungry you feel during the 12-hour recovery time. Plus, having the protein before bed didn't make you stronger or affect muscle damage markers compared to just having a non-caloric placebo [112].
But, having 20–40g of casein protein about half an hour before snoozing can help your body make more protein overnight, especially if you're a healthy adult guy wanting to build muscle after tough workouts. For older folks, the jury's still out on whether pre-sleep protein helps boost muscle mass and strength [113].
When it comes to how hard you push yourself during resistance training, it seems like whether you go to failure or not doesn't matter for sleep quality and heart rate variability. Both types of training impact sleep quality and autonomic modulation similarly throughout the night.
After a tough training session where you go all out, you might see a drop in your bench press by 7.2\% and half squat performance by 11.1\% the next day. But after an easier session, your performance might stay more stable. So listening to your body when deciding how intense your workouts should be is a crucial aspect [114].

A study found that having protein before bed can help muscles recover better after exercise. One group had a protein supplement comprising 27.5 g of protein and about half carbohydrates( compared to protein) every night, while another group had a placebo. Muscle strength increased more in the group of protein, compared to the placebo group(+164 and +130 kg, respectively) [115].
Lack of sleep can mess with hormones and muscle recovery. Not sleeping enough puts the body in a state that down muscle protein synthesis rates [116]. When people didn't sleep after intense exercise, they still recovered muscle strength. But it affected the body's response to inflammation and hormones. Even after going through muscle damage \& staying up for 48 hours straight, followed by a quick snooze of 12 hours, there was no impact on muscle recovery. It bumps up the blood IL-6 levels \& changes the hormone balance by boosting IGF-1 and cortisol while messing with the cortisol/total testosterone ratio [19].

Doing resistance exercises is good for getting stronger and healthier. However not sleeping well might affect how well resistance training works. Missing out on sleep might make it harder to lift heavier weights for some exercises, but not all. But if you keep on skimping on sleep night after night, your strength might take a hit when you're doing big moves involving multiple joints, like squats or bench presses. Single-joint seem to be less affected, though. Not getting enough shut-eye could mess with your overall muscle strength during compound movements unless the individual is super motivated or something like that. The impact of not sleeping enough on hormonal responses during resistance training isn’t clear [5].
Having a mix of carbohydrates and protein during and after a mix of resistance and endurance exercise didn’t change growth hormone or testosterone levels overnight, but it did increase cortisol levels in the body [117].
To make sure an individual gets enough sleep, they should have meals at the right times, and exercise, it's important for your metabolic health [116]. Research shows that having whey protein before bed can help your muscles grow stronger and bigger while you sleep. While people are interested in other kinds of protein like plant-based ones, we don't have proof that consuming them before bed works the same way(effectiveness).
It was evaluated that men who had 108 ± 0.02 g of protein per kg per day didn't fully recover after tough exercise done in the morning. Recovery didn't take place until the 3rd day had been completed. Moreover, irrespective of the source of protein, it did not facilitate muscle healing when hurtful eccentric exercise was enacted before noon [118]. Consuming 40g of protein before sleeping was seen to boost muscle protein synthesis during the night. This could be a good plan for older or sick individuals looking to keep their muscles healthy [119].
Another study discovered that just one night without any sleep made it harder for muscles to grow and led to more muscle breakdown. Lack of sleep caused a drop in MPS by almost 20\%. Cortisol (a stress hormone) went up by 21\% while testosterone (a muscle-building hormone) dropped by 24\% [120].

The lowered levels of MyopS in the sleep deficiency group could make sleeplessness effects worse on muscle mass and high-intensity interval exercise (HIIE) could be a useful solution. In an experimental observation, 24 young men were split into three: one got regular sleep with 8 hours in bed each night, another group had only 4 hours in bed, and the third group did the same short sleep but also did some high-intensity exercise.
It turns out, the muscle protein synthesis was way lower for the short-sleep group than for both the regular sleep and exercise groups. However, there was no change in any molecular markers used to indicate protein synthesis or degradation pathways typically activated by these pathways [23]. 
 Sleep deprivation in healthy young men was found to cause severe aggravation of glucose intolerance, alongside variable mitochondrial respiration rates in skeletal muscles. The scholars also emphasized the fact that their study had provided important information on possible mechanisms causing abnormality in glucose tolerance observed during sleep loss and it has been proposed that physical activity can be used as a therapy for alleviating negative consequences of such phenomenon [121].

\begin{figure}
    \centering
    \includegraphics[width=0.6\textwidth]{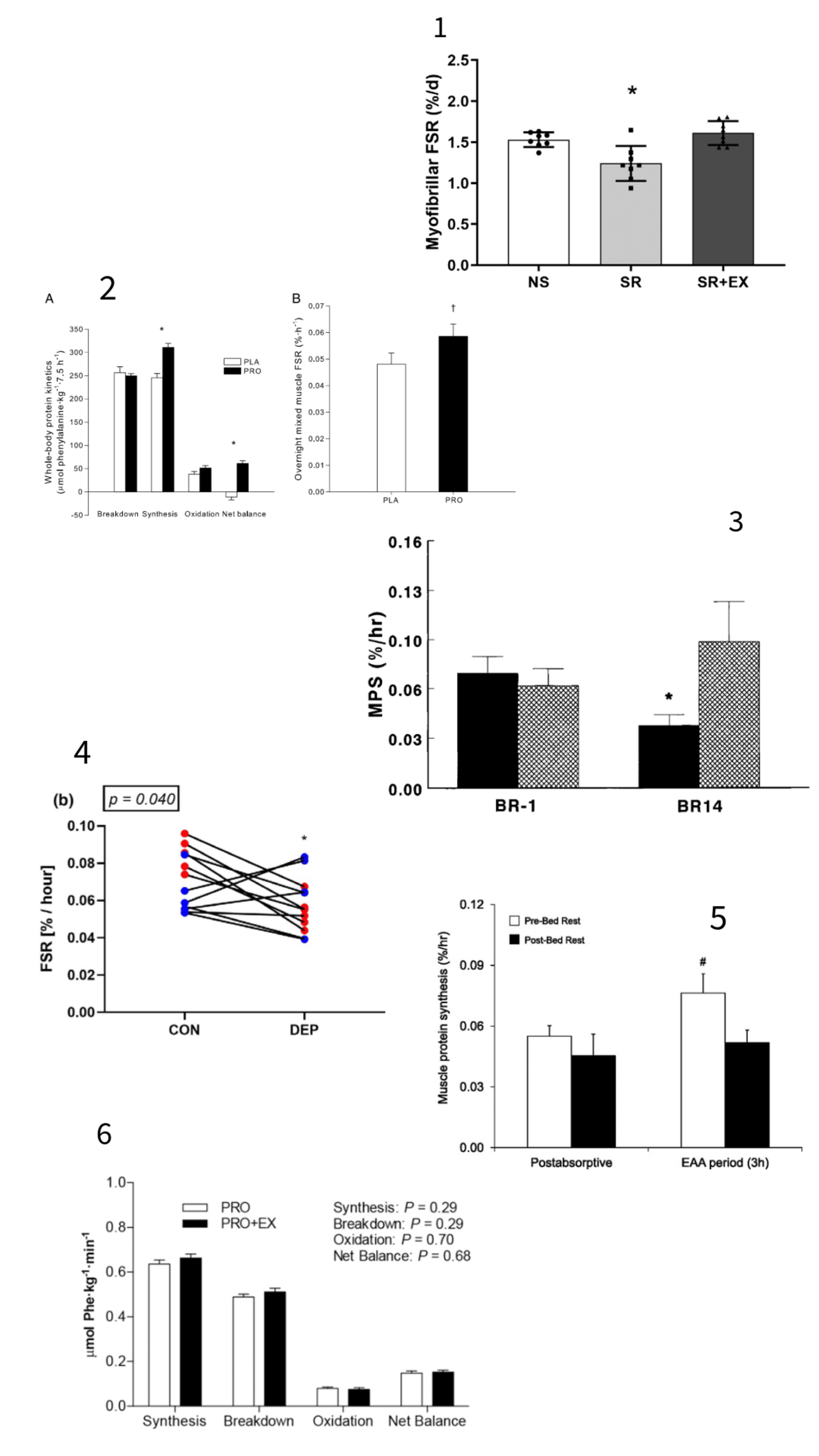}
    \caption{1. Evaluation of FSR at Normal Sleep,  Sleep Restriction, and their combination [23]. 
2. Examination of rates of multiple protein mechanisms in the PRO and PLA trial  [123].
3. MPS before and 2 weeks after bed rest in control subjects (solid bars) and subjects involved in resistance program (crosshatched bars) [53].
4. Postprandial mixed muscle FSR measured in the control and sleep-deficient conditions. The red dots identify male subjects and the blue ones depict females [120]. 
5. Mixed MPS in muscles of older adults in the post-absorptive state and throughout the EAA devouring state [127].
6. Various rates of protein mechanisms in older men after protein ingestion and protein with exercise [122].
}
    \label{fig:enter-label}
\end{figure}

The study conducted by [122] appraised if physical activity in the evening could boost overnight muscle protein synthesis when paired with pre-sleep protein intake among older men. This study observed a 31\% and 27\% increase in overnight myofibrillar protein synthesis in participants who combined protein intake with exercise compared to those who only had protein. Moreover, the body effectively processes and absorbs the consumed protein before sleep, leading to a surge of amino acids circulating during the night. 
As reported in another study with sixteen young men, those who consumed protein before bedtime experienced a 22\% higher mixed muscle protein synthesis rate than those who didn't.
The conclusion drawn from these studies was that consuming dietary protein right before sleep enhances the availability of plasma amino acids, thus promoting muscle protein synthesis, which aids in improving overall protein balance throughout the night [123].
 Additionally, delivering protein during sleep can elevate plasma amino acid levels and activate muscle protein synthesis, contributing to increased protein balance while sleeping.
Interestingly, providing high doses of dietary protein (40g) during sleep was well-tolerated by participants without causing any digestive issues or discomfort. The research showed that there was no interruption to participants' sleep patterns after ingesting the provided proteins or placebos. It is vital to note that the fractional synthesis rates were notably higher when individuals consumed protein right before sleep compared to the placebo group which indicates abundant integration of amino acids utilized through protein( essentially labeled) [124].

On a different note, a twelve-week program of resistance training coupled with post-workout and pre-sleep protein supplementation did not significantly enhance muscle mass or strength gains in active older men. Muscle fiber hypertrophy was observed in both groups; however, there were no significant differences seen among them. The meal comprised either protein (21 g) or an energy-matched drink(25 g carbohydrate) after the session and each night before sleep. The type II muscle fibers experienced hypertrophy of the muscle fiber [125].
Ten aged individuals between 65 and 80 lifted weights for a week followed by lying in bed for five days. Their leg muscles grew more with exercise (1.76 vs. 1.36). Muscles shrank during bed rest but less in the exercised leg (1.07 vs. 1.30). Postprandial aMyoPS (acute postabsorptive and postprandial myofibrillar protein synthesis) rates augmented above postabsorptive values in the leg that went through training. Despite short-term resistance did not counterbalance the relative reduction in iMyoPS and muscle mass during rest, the blood sugar and cholesterol stayed the same during the resting period [126].
 Essential amino acids are vital for our muscles. These amino acids help with signaling in our bodies, promote muscle protein synthesis, and maintain muscle mass. Likewise, they stimulate the amino acid transporter, mTORC1 signaling. It's surprising to note that we don't have much info on how inactivity affects this process despite it being crucial for muscle mass maintenance.
 A study involving six older adults showed that just 7 days of bed rest led to a 4\% decrease in leg lean mass. Despite this, levels of a key protein related to muscle growth like postabsorptive mTOR increased, whereas MPS of post-absorptive condition was similar. This change in response to amino acids might contribute to faster muscle loss when we're not active [127].
Moving on to sleep - it is understood that not getting enough sleep can cause our muscles to shrink. This happens because our bodies produce more of certain hormones that break down proteins (glucocorticoids) and less of others that help with muscle growth(testosterone, growth hormone). Resistance exercise can be super helpful for keeping muscles healthy for everyone. This will help your body grow stronger because it boosts growth hormones like testosterone and IGF1 (Insulin-like growth factor 1), MPS by activating mTOR [128].
When athletes don't get enough sleep, they're more likely to get hurt because their bodies break down protein faster and don't build muscle as well [22].

Moreover, after not moving around for a while or due to inactivity, our muscles can become resistant to the effects of food. Even if we eat well, we might still lose muscle mass over time. A study found that after just 7 days of eating a regular diet and then switching to simulated weightlessness for 14 days, people experienced a significant decrease in MPS of a staggering 50\%, in both leg and overall lean body mass.
Likewise, there was a 46\% reduction in fractional MPS by tracer integration into muscle protein. Not moving around makes you lose muscle protein which affects your whole body and muscles [129].
It is observed that not getting enough sleep can indeed screw up your memory and thinking skills. Plus, it increases the chances of getting sick and can even mess with your genes and brain structure. On top of that, it heightens the risk and progression of many diseases. Alters molecular signaling and gene expression and changes dendrites in neurons [130].
Sleep is very important for keeping your body in check - like sure your bones, fat, and muscles stay healthy. Indeed, the data collected proved that processes led by sleep play a massive role in the regulation of endocrine functions which are important in tissue repair renewal and growth of the body [131].
Getting more sleep could genuinely help you heal faster from muscle injuries caused by working out. It helps boost certain proteins and reduces inflammation. Although research on sleep and its relation to sports is still in its inception phase, various theoretical clues regulating the scientific articles suggest various mechanisms of sleep-exercise muscle injury [132].

In a recent study, researchers looked at how different amounts of sleep affected genes and inflammation in young guys. Some got a full 8 hours of sleep each night, some only 4 hours( restricted from sleep: SR), and others did intense workouts(EX) after getting just 4 hours of shuteye. After the intervention, the SR group had more inflammation and immune response activities. However, the SR+EX group had fewer of these activities. Genes affected by lack of sleep also showed less activity in the SR group [21].
Another study found that men with low muscle mass were more likely to have trouble sleeping well. Women who didn't get enough quality sleep had weaker grip strength and lower muscle mass too. The link between good sleep and muscle health was studied in 1196 people. Over half were women, around 68 years old. It found that 19.1\% of women and 13.4\% of men had bad sleep. Men with low muscle mass often had very bad sleep (9.1\% vs 4.8\% in women) [133].
A trial examined how taking a break in bed for 10 days affects the muscles of older folks. The study found that during this time, there was a noticeable decrease in muscle mass, especially in the legs. Older participants lost more lean tissue in 10 days than younger ones did in 28 days. However, their decrease in protein synthesis and strength was similar to the younger group after 14 days. Interestingly, even before bed rest, these adults were already lacking nitrogen balance despite eating enough protein [134].
Recovering after exercising is mandatory for the body to bounce back. It's all bringing things back to normal after a workout - replacing lost fluids and nutrients, fixing up damaged tissues, and regulating body temperature and heart function. These things should all happen in good time before the next workout or event takes place [135].

A recent trial showed that using neuromuscular electrical stimulation and protein supplements (NMES+PRO) helped protect muscle mass and function in older adults after just 5 days of bed rest. The NMES+PRO group got muscle zaps to their thigh muscles (40 minutes per time, three times a day) and then drank a 17g protein shake. The other group, called CON (Control), drank a similar drink but without protein. Participants who received NMES to the quadriceps along with protein supplements managed to maintain thigh muscle mass but not muscle function during this time [136].
 When older folks did Resistance Training, their arousal index  went down from 22.27 ± 11 events/h to 20.41 ± 8.57 events/h (t = 2.10 and p = 0.04). Throughout stage 1 sleep, the control group (those who didn't do RT) had a slight increase from 4.96\% at the start to 5.40\%. But get this, the resistance group saw a decrease from 8.32 to 6.21\% after their workout session.
The control group had more awakenings per hour compared to the resistance group post-exercise session. Also, those in the control group spent more time in stage-1 sleep than those in the resistance group.
However, after an intense exercise session, those in the resistance group took less time to get into REM (Rapid Eye Movement) sleep compared to their counterparts in the control group [137]. 
Doing resistance exercises at any time of the day can help improve sleep quality compared to not doing resistance exercises. Resistance exercise can provide great benefits for falling asleep and staying asleep, especially for individuals with osteoporosis, sarcopenia, anxiety, or depression. While the timing of resistance exercise doesn't change average nocturnal blood pressure or different types of sleep it does show that doing resistance exercise at any time of day can help with staying asleep more than skipping this type of workout [138]. 
If you do intense resistance exercises for two days in a row, it can reduce activity in the parasympathetic nervous system during sleep. One study had participants do lower-body exercises in the morning and upper-body exercises in the afternoon. During the sleep tests, there was a big drop in stage 2 sleep on the second exercise day. The high-frequency heart rate fluctuations were significantly lower at night 2 compared to other nights [139].
When moderately intense aerobic and resistance exercises are done in the evening, it doesn't seem to affect sleep quality at night. Core body temperature increases during these exercises but goes back to normal before bedtime. Doing moderate-intensity resistance exercises 90 minutes before bed doesn't disrupt sleep for healthy young men either. Exercise also doesn't impact alertness, physical tension before sleep, or how well someone feels they slept [140].

One study found that folks didn't sleep longer after exercising but felt their sleep quality improved. Exercise can make you sleep better, as shown by lower scores on the Pittsburgh Sleep Quality Index \& better subjective sleep quality, less time drifting off, \& fewer sleep medications. However, actual time spent sleeping, disturbances, \& efficiency weren't significantly affected. Bad sleep can mess with how well you think and move, especially during hard or long physical activities [141]. Athletes often struggle to manage their sleep, hurting their thinking, effort perception, and workout health. To boost physical performance, we need to understand the body's responses to lack of sleep better [142].

Working out with weights after drinking different amounts of whey protein, casein protein, or a placebo 30 minutes before bed didn't change appetite much over time. Cortisol levels went up after lifting weights. Before lifting weights, feeling tense was linked to being hungry and wanting to eat a good amount. Afterward, cortisol was related to how you felt and how much energy you had. The protein amounts before bedtime didn't affect morning hunger and cortisol levels but did connect cortisol levels with mood and appetite [143].

\begin{center}

Findings from multiple studies

\end{center}
\begin{tabular}  {|p{1cm}|p{2.5cm}|p{2.5cm}|p{4cm}|p{4cm}|}

\hline
Authors & Aim of Study & Methods, Subjects & Results & Notes \\
\hline
[77] & To determine whether 40gm whey stimulates larger MPS than 20gm & 30 healthy RT trained, with LBM 65 to 70 kg & 
MPS greatly stimulated to ingestion of 40 g (0.059 ± 0.020\%/h ) compared with 20 g (0.049 ±  0.020\%/h) of protein & Individuals with greater LBM would require more protein for greater stimulation of MPS during recovery from whole-body RE compared with individuals with lower LBM. \\ \hline
[78] & To examine dose-response relation after ingestion of 0, 10, 20, 40g whey 10 mins after RE & 48 RT trained men, (8*10 leg presses and leg extensions at 80\% 1RM) & 
MyoMPS ↑ above 0g
protein by 49\% and 56\% with the ingestion of 20 and 40g whey protein, urea production ↑ with the ingestion of 40g & 
20g of whey is enough for the max stimulation of postabsorptive rates of Myo MPS in 80-kg resistance-trained, young men. \\ \hline
[92] & 
To examine the varying stimulation of contractile myofibrillar proteins following the ingestion of 25g of protein and assess how it is influenced by resistance training (RT).
& 7 active males, (26 ± 3 years; 84.8 ± 4.5 kg),unilateral leg exercise & 
 MyoMPS in the exercised leg was stimulated above FAST at 1, 3, and 5 h (100, 216, and 229\%), increase at 5h was greater than the FED state 
& 
Doing resistance training (RT) right after eating protein helps build up the muscle protein quickly.
\\ \hline

[79] & 
To evaluate ingested protein dose response of  0, 5, 10, 20, or 40g whole egg protein & 
6 healthy young men, intense leg-based RT & Your body uses protein best when you eat about 20g of it, and Leucine oxidation increases a lot when you have 20 or 40g of protein. & Excess dietary protein stimulates irreversible oxidation. \\ \hline
[80] & 
To identify how much protein can be used in a single meal
& Literature review & MPS is maximized in young adults with an intake of 20-25g of high-quality protein & Consuming protein at an intake of 0.4 g/kg/meal across four meals to reach a minimum of 1.6 g/kg/day could maximize anabolism. \\ \hline
[81] & 
To evaluate MPS response to RE and protein ingestion during energy deficit & 
15 participants, MPS during EB and 5 days ED, MPS in ED after acute RE & 
MPS were 27\% ↓ in ED than EB but RE stimulated MPS to rates equal to EB. Ingestion of 15 along with 30 g of protein after REX in ED ↑ MPS 16 and 34\% above resting EB & 
Combining RE with ↑ protein availability after exercise ↑ rates of skeletal MPS during short-term ED and preserves muscle mass in long term. \\ \hline
[82] & 
To investigate the impact of particular ingestion patterns on muscle protein synthesis (MPS) over a 12-hour duration. & 24 healthy trained men, leg extension,
4 sets of 10 repetitions & 20g whey consumed every 3 hours is superior to BOLUS or PULSE & To achieve peak muscle mass, the protein distribution is a key component. \\ \hline

[86] & 
To identify effects in MPS after whey and caseinate intake post-RT & 24 elderly men and women & 
No difference in MYO and collagen FSR compared to whey and casein & MPS are similar with caseinate ingestion before and after exercise. \\ \hline
[61] & 
To examine effects of RE volume on MyoMPS in young men & 8 RT trained men, 24 ± 5 years, unilateral leg extension exercise at 70\% 1RM, 1 or 3 sets & 
 When we do 3 sets of resistance exercises, muscle protein synthesis goes up more than only doing 1 set. This increase lasts longer too, lasting up until 29 hours post-exercise.  & Basically, doing more sets seems to help build up muscle protein better over time. \\ \hline

\end{tabular}

\begin{center}
\begin{tabular}  {|p{1cm}|p{2.5cm}|p{2.5cm}|p{4cm}|p{4cm}|}
\hline
Authors & Aim of Study & Methods, Subjects & Results & Notes \\
\hline

[63] & To know how long your muscles work during low-intensity resistance exercise affects making specific types of muscle protein. & There were 8 guys, an average age of 24, to either do exercises until failure that took either 6 seconds (SLOW) or just 1 second (CTL). & After a day or so of recovery, it turned out that the guys who did the slower reps had higher muscle protein synthesis than the others. & Substantial muscle time in tension delayed the stimulation of MyoPS 24–30 h after resistance training session \\ \hline
[68] & To evaluate whether low-intensity RE training combined with blood flow restriction (REFR) increases muscle size
& 6 young male,
Bilateral leg extension at 20\% 1RM with pressure cuff(REFR), without cuff(CTRL) & 46\% increase in MPS following REFR & Activation of the mTOR signaling pathway seems to be an important cellular mechanism that would help to promote the enhanced MPS during REFR. \\ \hline
[72] & To determine how RE, HIIT affected the integrated day-to-day response of MPS & 22 sedentary men,  67±4 years, & Myofibrillar protein FSR was elevated, relative to rest, at 24 and 48 hours following RE and HIIT & HIIT largely ↑ myofibrillar and sarcoplasmic fractional synthetic rate in older groups \\ \hline

[76] & To identify how resting between sets during resistance exercise impacts muscle growth & They got 16 men to do leg exercises at different rest times - either just 1 minute or a full 5 minutes - at certain weights. & The rate of making muscle protein went way up right after exercise for both rest intervals - whether they rested for just a minute or five.  & 
Short rest intervals of one minute between sets in RE were associated with ↓ MyoPS in the early recovery phase after exercise, compared to longer rest intervals of five minutes. This ↓ may be due to impaired activation of intracellular signaling pathways. \\ \hline

[52] & They looked into how resistance training affects your body's response to eating and making mixed muscle protein. & 
10 young men, knee extension exercise & After some resistance training, both legs showed an increase in muscle protein creation at around 4 hours post-exercise & 
RT reduces the protein synthetic response to acute resistance exercise, despite the heightened early rise in FSR \\ \hline
[65] & To evaluate whether fasted state skeletal MPS after RE is altered with training & 8 wk, knee extension, 80\% 1RM, & 
After RT, resting mixed MPS rate was lifted by 48\%, MyoMPS was invariable at rest following training & Strength training had a positive impact on the immediate response of muscle protein synthesis (MPS) to resistance exercise (RE) by slowing down the increase in the production non-myofibrillar proteins. \\ \hline
[53] & 
To examine if RE maintains skeletal MPS during bed rest & 12 healthy men,2 wks, one group performed RE and BR(BREx), whereas the other did strict bed rest (BR) & 
The group that did moderate RE didn't experience any changes in MPS and had significantly higher levels compared to the other group. & 
When people engage in moderate resistance exercise, it can help combat the decline in MPS that usually happens during periods of bed rest. \\ \hline

[55] & 
To identify the effects of 3 months of RT in MyoMPS in young and old subjects & 18 subjects & 
 Older individuals did show a 27\% slower rate than the young ones & In healthy older subjects, the MyoMPS process in the vastus lateralis muscle is slower than in young adults. \\ \hline

\end{tabular}
\end{center}

\begin{center}
\begin{tabular}  {|p{1cm}|p{2.5cm}|p{2.5cm}|p{4cm}|p{4cm}|}
\hline
Authors & Aim of Study & Methods, Subjects & Results & Notes \\
\hline
[43] & To determine whether muscle hypertrophy is the outcome of RT & 10 men, 1,3,10 week program &  The changes in MPS following an initial bout of resistance exercise during the first week of training were more significant compared to those seen after exercise sessions during the third and tenth weeks. &  Muscle hypertrophy results from periodic increases in MyoPS, primarily after minimizing muscle damage over time. \\ \hline

[46] & 
The aim was to investigate potential differences related to age in how MPS responds to resistance exercise among young and older men. & 2 groups, 25 each, old(70 ± 5 years) and young (24 ± 6 years) & There seems to be a strong link between post-exercise MPS levels at 1-2 hours and the intensity of the training being performed by each group.
 & 
Aged men showcase anabolic resistance of MPS to RE \\ \hline
[44] & To evaluate the time course for elevated MPS after RE & 
6 healthy young men, 12 sets, 6-12 elbow flexion exercises & By 24 hours post-exercise, MPS levels nearly double but then decrease rapidly so that by 36 hours they have mostly returned to their starting point. & 
MPS is raised at 4, 24, and 36 hrs by 50, 109, and 14\% respectively, following a session of heavy RT. \\ \hline
[94] & To determine the result of protein ingestion combined with RE on muscle size and strength & 
22 young men, 14-week program, PRO or CHO fed after RT & PRO group showed hypertrophy of type I (18\%±  5\%) and type II (26\% ± 5\%;) muscle fibers & 
A slight upper hand of protein supplementation over CHO  throughout RT on mechanical muscle function was observed. \\ \hline
[83] & 
To evaluate whether rapid aminoacidemia enhances myofibrillar protein synthesis after RE & 8 men, BOLUS 25g or 10 2.5g drink(PULSE) every 10 min & MPS was ↑  to a larger degree after bolus the following pulse initially (1–3h: 95\% contrast to 42\%) and later (3–5h: 193\% contrast to 121\%) & The intake of amino acids directly following a training session has been shown to boost MPS and promote anabolic signaling more effectively than consuming an equivalent amount of protein in smaller doses throughout the day. \\ \hline
[88] & 
To identify Leucine’s and EAA’s role in MyoMPS at rest and after RE &   
24 males, unilateral RE, 6.25g whey with leucine equivalent to 25g whey, 6.25g whey with no leucine only & Specifically, whey protein remained significantly higher than other forms like leucine or essential amino acids-leucine after resistance exercise. &  Adding a small amount of whey protein alongside leucine or other essential amino acids proved just as effective as consuming a complete protein. \\ \hline
[90] & 
To determine MyoMPS after soy protein ingestion & 
30 elderly, 71 ± 5, 0,20 or 40 g soy after RE & Select studies have indicated that lower amounts of protein are not as effective for increasing MPS compared to larger doses following resistance exercise. & 
 Whey showed superior abilities in boosting MPS levels both at rest and after intense exercise sessions. \\ \hline
[112] & 
To determine if pre-sleep protein enrichment after a single session of evening RE improves next-day ability or recovery & 18 RT-trained men, 40g casein PRO or PLA 30 min before sleep & There weren't any notable differences observed between different groups regarding physical performance  & 
 There weren't any notable differences observed between markers of muscle damage when comparing pre-sleep consumption of protein supplements with placebos among weight-trained men. \\ \hline

\end{tabular}
\end{center}

\begin{center}
\begin{tabular}  {|p{1cm}|p{2.5cm}|p{2.5cm}|p{4cm}|p{4cm}|}
\hline
Authors & Aim of Study & Methods, Subjects & Results & Notes \\
\hline
[114] & 
To examine whether RT intensity affects sleep quality and strength &
 15 RT trained athletes, 2 RT sessions leading to failure and non-failure, 90s rest, 75\% 1RM & 
 No variations were observed between failure and non-failure exercise bouts on SQ & The test showed that one workout session leading to failure causes tiredness that lowers the maximum weight in bench press by 7.2\%, half squat by 11.1\% the next day. \\ \hline
 
[115] & To examine whether sleep after PRO consumption augments muscle mass and strength  & 44 men (22 ± 1)y, 12wk, 27.5g PRO or PLA before sleep & 
Muscle strength increased in the PRO group than in PLA & 
Protein ingestion before sleep signifies an efficient tactic for ↑ strength and mass throughout RE in young men \\ \hline
[113] & 
To identify the effects of Pre-sleep protein on MPS & 
Literature Review & 
Eating 20–40 grams of casein about 30 minutes before bedtime boosts protein synthesis in the whole body during the night for young men. & 
Pre-sleep PRO consumption can augment the muscle adaptive response during 10–12 weeks of RE in young \\ \hline

[118] & To determine effects of pre-sleep plant-based vs whey PRO on muscle recovery after effective morning RE & 
27 middle-aged men, 5 sets of 15, knee extension, 40g whey, PLA or rice and pea combination & 
No significant differences between groups for any outcome measure. & 
Having 1.08 ± 0.02 grams per kilogram per day of protein didn't help recover from tough eccentric exercise at +72 hours, and having protein before sleep, regardless of the source, didn't change this result. \\ \hline

[119] & To evaluate whether PRO ingestion before sleep increases overnight MPS & 48 men(72 ± 1 y), 20 and 40g casein, 20g casein + 1.5g LEU or PLA before sleep & 40 g protein before sleep ↑ myofibrillar protein synthesis rates during overnight sleep & Protein ingested before sleep is properly digested and absorbed throughout the night  \\ \hline
[129] & 
To examine whether prolonged bed rest decreases skeletal muscle and whole body MPS & 6 male, 14 days Bed Rest, & MPS ↓ by 50\%, whole body protein synthesis ↓ by 14\% & 
The loss of body protein with inactivity is predominantly due to an ↓  in MPS and this decrease is reflected in both whole body and skeletal muscle measure \\ \hline
[120] & 
To evaluate the effect of acute sleep deficiency on skeletal MPS and the hormonal condition & 
7 males and 6 females, 1-night total sleep deprivation and normal sleep, randomized crossover design & Sleep deprivation decreased MPS by 18\%, plasma cortisol increased by 18\% & Just one night without sleep can make your resistant to muscle building and more likely to break down muscle. \\ \hline

\end{tabular}
\end{center}

\begin{center}
\begin{tabular}  {|p{1cm}|p{2.5cm}|p{2.5cm}|p{4cm}|p{4cm}|}
\hline
Authors & Aim of Study & Methods, Subjects & Results & Notes \\
\hline
[23] & To determine the effect of sleep restriction, with or without HIIT on MPS & 
24 healthy young men, 2 nights controlled baseline sleep and 5 night intervention period,3 groups, 8hr(SR), 4hr,4hr+EX & MyoPS was lower in the SR group with FSR  1.24 ± 0.21 (\%/day) & 
MyoPS is acutely ↓ by sleep restriction, although MyoPS can be maintained by performing HIIE. \\ \hline
[124] & 
To evaluate the role of intragastric protein administration on overnight MPS
 & 
16 healthy elderly men, casein PRO or PLA during sleep & 
Overnight muscle protein FSR was much larger in the PRO experiment &  Dietary protein administration during sleep stimulates MPS and improves overnight whole body protein balance \\ \hline
[126] & To see how short-term exercise before bed rest affects muscle protein synthesis and loss.

 & 
10 men, 65-80 yrs, 4 bouts of unilateral RT, 7 days prior to 5-day BR & 
Muscle protein synthesis rates were higher in the exercised leg compared to the non-exercised leg during prehabilitation. & Short-term resistance training before bed rest increased muscle protein synthesis rates in older men but couldn't prevent the decrease in muscle protein synthesis and muscle mass during bed rest. \\ \hline

[73] & 
To evaluate whether RE stimulates MPS in lean and obese adults & 9 lean, 8 obese, unilateral leg exercise, 19-32yrs & RE enhanced MPS (50\% growth) in both groups (P < 0.001), with no difference between lean and obese. & 
Obese young adults had a similar rise in MPS to that of their lean mates. \\ \hline
\end{tabular}
\end{center}

\textit{REX, resistance exercise, RE, resistance exercise, EB, energy balance, ED, energy deficit, MyoMPS, myofibrillar muscle protein synthesis, RT, resistance training, MPS, muscle protein synthesis, Myo, myofibrillar, FSR, fractional synthesis rate, REFR, resistance exercise flow restriction, CTL, control, CTRL, control, mTOR, mammalian target of rapamycin, HIIT, high-intensity interval training, MyoPS, myofibrillar protein synthesis, BR, bed rest, BREx, bed rest after exercise, CHO, carbohydrates, PRO, protein, PLA, placebo, iMyoPS, integrated myofibrillar protein synthesis, S20, 20 grams soy, W20, 20 grams whey, EX, exercise, SQ, sleep quality}

\section{DISCUSSION}
After evaluating multiple studies, it has been pointed out that limited research has been conducted on the combined influence of resistance exercise, protein supplementation, and sleep/recovery. Whereas, there have been hundreds of individual studies on each of them. This review has illustrated multiple training protocols as well as the timing, quality, and quantity of dietary protein requirements for optimal MPS. The role of sleep and its effects on exercise intensity, strength, protein digestion, and, of course, whole-body MPS in a given period have been established. 

For having a good quality of life, food and adequate sleep are very essential. Therefore, having proper nutrition plays a significant role in our day-to-day lives. Protein is required for preserving and growing muscle mass. Likewise, having adequate sleep is mandatory, as an average human spends one-third of their lifetime sleeping. The rise in population gives an invitation to a rapid boost of new kinds of diseases and climatic changes. So, in the coming years, it will become a challenge to prevent diseases and keep a larger fraction of the population physically fit and active. Adding to that, understanding the mechanisms involved in MPS through the synergy of these three aspects becomes essential. 

New studies and research would bring new potential for preserving muscle mass. For example, new studies could benefit elderly individuals. Large muscle mass provides strength and energy to perform more physical activities. Therefore, dependency on younger adults for developmental work could be slightly minimized. Hence, extensive research needs to be carried out in this field to provide better lifestyles and more active individuals. 

\section{CONCLUSION:
}
This review has examined the potential roles of resistance training methodologies, protein dosage, supplementation, and other critical parts of sleep and recovery in influencing muscle protein synthesis. Resistance exercise, on the one hand, provides a strong stimulus to muscle growth, while protein supplementation provides the necessary amino for muscle repair and growth. Of these three factors, sleep has been recognized as one of the important elements sustaining physiological and biochemical processes integral to muscle recovery and growth. 
 
It has been demonstrated that a dose of 20-30 grams of protein at each meal will induce a maximal MPS response, provided the frequency is three to four times daily [77]. Resistance exercise combined with increased protein provision after exercise has been shown to enhance skeletal MPS rates during acute ED and can also contribute to preserving muscle mass during longer-term ED. [81].  30 grams of PRO can increase MyoPS, it shoots up to a huge 46\% per hour compared to no protein at all. That is one leap! But on an interesting note, while it does wonders after endurance exercise, it doesn't do much for mitochondrial protein synthesis [85]. Whey protein dominates when it comes to pumping muscle protein synthesis. It outs soy, intact casein, and even hydrolyzed casein [90]. After having chugged down some whey, the muscle protein synthesis goes up by a whopping 93\% more than that by casein and by 18\% more than soy. And get this - after exercise, the muscle protein synthesis post-whey intake spikes up by 31\% higher than soy and an incredible 122\% higher than casein [98].
Now, moving to EAA CHO consumption after resistance exercise: it does a great job in promoting muscle protein synthesis during recovery post-exercise. However, here comes the catch: it doesn't do anything to help boost muscle protein synthesis before a workout [99]. In terms of one year of protein supplementation versus an isocaloric control supplement, they do not show any significant effect on MPS levels either at the pre- or post-meal stage or during the recovery period [107]. More recently, one study has emphasized that high-protein (leucine-enriched whey) consumption may prevent—or, at least, partly offset—the negative effects of weight loss on skeletal muscle mass [109].  It is confirmed that the human muscle synthetic rate increased by as much as 50\% as early as 4 hours after heavy resistance training, before reaching as high as 109\% at the 24-hour mark post-workout [45]. A mixed muscle protein synthesis rate is induced within a time  as short as 4 hours following an intensive plyometric muscle contractions-based exercise. Curiously, resistance training somewhat puts a damper on this spike in rates [47].

At the start of a workout, both older adults and younger individuals experienced significant improvements in their leg muscles, with no noticeable change in overall muscle breakdown [49]. Over three days, the leg that was exercised showed a marked increase in muscle growth compared to the untrained leg[ 51]. As expected, the trained leg exhibited a much greater increase than the untrained leg within four hours; however, it was also anticipated that both legs' muscle protein synthesis rates would rise after eight weeks of one-legged resistance training [52]. Additionally, during both concentric and eccentric resistance exercises, muscle protein synthesis was elevated above resting levels at every measured time point: 112\% at three hours, 65\% at 24 hours, and finally 34\% at 48 hours. Moreover, the rate of muscle breakdown was also heightened after exercise at three hours and again at 24 hours, but by the end of that period, it returned to resting levels [56].
Thus, for beginners, the muscle protein synthesis immediately following that first workout doesn't necessarily correlate with muscle growth from subsequent resistance training [57]. Now, let's turn our attention to myosin heavy chain and mixed protein synthesis rates in older adults. It appears that these rates decline more significantly compared to actin, and the variability in measurements is greater for actin than for myosin heavy chain or mixed muscle protein [60].
When you engage in exercise and your muscles contract, muscle protein synthesis increases. This is true for both men and women, with no significant differences in leg muscle responses among young adults [45]. Additionally, muscle protein synthesis returns to near baseline levels within a few hours after exercising for both younger individuals (around 24 years old) and older adults (approximately 70 years old) with similar body mass indexes [46].

Performing quad extension exercises at 0.3 times your maximum potential, with a slow lifting pace (taking 6 seconds to lift and lower), can significantly enhance muscle protein synthesis compared to executing the same movement quickly (1 second up and down). This indicates that the duration your muscles are under tension during workouts may be crucial for promoting muscle growth [63]. 
Endurance exercise (EE) has been shown to increase mitochondrial protein synthesis in both untrained individuals (by an impressive 154\%) and trained individuals (by 105\%), but it did not have the same effect on myofibrillar synthesis after 10 weeks. Essentially, performing endurance exercises with one leg does not immediately stimulate myofibrillar muscle protein synthesis, regardless of training status. It primarily seems to enhance mitochondrial protein synthesis [64]. 
Additionally, research indicated that increases in p70S6K phosphorylation and alterations in muscle androgen receptor (AR) protein content were associated with muscle hypertrophy. This implies that the internal processes within your muscles play a more significant role in achieving those gains than overall bodily functions [76]. 
Regarding resistance training, pushing yourself to failure does not appear to significantly affect sleep quality compared to stopping just before that point [114]. Moreover, consuming protein before bedtime did not seem to influence muscle damage markers or strength levels the following morning for regular weightlifters. However, having some protein before sleep might help reduce morning hunger [112].

Eating protein before bed supports muscle growth during sleep. Those who consumed protein experienced a greater increase in muscle strength compared to those who did not [115]. Even if you don’t get enough sleep after exercising, your muscles can still recover, although it may affect how your body reacts to inflammation and hormones [19].
Consuming around 40 grams of protein before bed aids your muscles in synthesizing more protein while you rest [119]. However, missing out on sleep for just one night can disrupt your body’s muscle-building capabilities and may lead to a catabolic state [120]. 
Having protein before sleep increases amino acids in your bloodstream, promoting muscle growth while maintaining your body’s overall protein levels. Plus, you don’t need to worry about feeling uncomfortable; a reasonable amount of protein before bed won’t upset your stomach [124].
Athletes who don’t get enough sleep are at a higher risk of injury. This is due to their bodies signaling more breakdown of muscle proteins and less synthesis, which weakens their muscles over time [22].
When muscle protein synthesis decreases, it typically indicates that fewer amino acids are being transported within the cells. This suggests that during periods of inactivity, our muscles begin to lose proteins, impacting both our overall body and specific muscle groups [129].
In a study involving older adults on bed rest for 10 days, they experienced significant muscle mass loss, particularly in their legs. Interestingly, even before their bed rest, they were not achieving a sufficient nitrogen balance despite consuming the recommended amount of protein [134].
Finally, many athletes face challenges with sleep management, which can negatively affect their cognitive abilities, effort levels, and overall health related to exercise adaptation [142].

In summary, the integration of proper resistance training, adequate sleep, and protein supplementation forms a strong trio for achieving optimal muscle protein synthesis for individuals of various age groups. The journey of understanding the best practices for muscle protein synthesis is ongoing, and this review offers effective strategies for improving health and physical performance to all health practitioners and individuals. As this field continues to develop and generate more fruitful findings, upcoming studies need to understand the dynamics among them in greater detail to establish a foundation for the development of more effective and customized guidelines that meet the needs of individuals.

\newpage
\bibliographystyle{unsrtnat}
\bibliography{references}  

[1]	B. S. Gordon, A. R. Kelleher, and S. R. Kimball, “Regulation of muscle protein synthesis and the effects of catabolic states.,” Int. J. Biochem. Cell Biol., vol. 45, no. 10, pp. 2147–2157, Oct. 2013, doi: 10.1016/j.biocel.2013.05.039.

[2]	R. R. Wolfe, “The role of dietary protein in optimizing muscle mass, function and health outcomes in older individuals,” Br. J. Nutr., vol. 108, no. S2, pp. S88–S93, Aug. 2012, doi: 10.1017/S0007114512002590.

[3]	P. J. Atherton and K. Smith, “Muscle protein synthesis in response to nutrition and exercise,” J. Physiol., vol. 590, no. 5, pp. 1049–1057, Mar. 2012, doi: 10.1113/jphysiol.2011.225003.

[4]	B. E. Phillips, D. S. Hill, and P. J. Atherton, “Regulation of muscle protein synthesis in humans,” Curr. Opin. Clin. Nutr. Metab. Care, vol. 15, no. 1, pp. 58–63, Jan. 2012, doi: 10.1097/MCO.0b013e32834d19bc.

[5]	O. E. Knowles, E. J. Drinkwater, C. S. Urwin, S. Lamon, and B. Aisbett, “Inadequate sleep and muscle strength: Implications for resistance training,” J. Sci. Med. Sport, vol. 21, no. 9, pp. 959–968, Sep. 2018, doi: 10.1016/j.jsams.2018.01.012.

[6]	W. J. Kraemer and N. A. Ratamess, “Hormonal Responses and Adaptations to Resistance Exercise and Training,” Sports Med., vol. 35, no. 4, pp. 339–361, 2005, doi: 10.2165/00007256-200535040-00004.

[7]	R. C. CASSILHAS et al., “The Impact of Resistance Exercise on the Cognitive Function of the Elderly,” Med. Sci. Sports Exerc., vol. 39, no. 8, pp. 1401–1407, Aug. 2007, doi: 10.1249/mss.0b013e318060111f.

[8]	R. W. Braith and D. T. Beck, “Resistance exercise: training adaptations and developing a safe exercise prescription,” Heart Fail. Rev., vol. 13, no. 1, pp. 69–79, Feb. 2008, doi: 10.1007/s10741-007-9055-9.

[9]	A. R. Hong and S. W. Kim, “Effects of Resistance Exercise on Bone Health,” Endocrinol. Metab., vol. 33, no. 4, p. 435, 2018, doi: 10.3803/EnM.2018.33.4.435.

[10]	K. F. Koltyn and R. W. Arbogast, “Perception of pain after resistance exercise,” Br. J. Sports Med., vol. 32, no. 1, pp. 20–24, Mar. 1998, doi: 10.1136/bjsm.32.1.20.

[11]	M. Toigo and U. Boutellier, “New fundamental resistance exercise determinants of molecular and cellular muscle adaptations,” Eur. J. Appl. Physiol., vol. 97, no. 6, pp. 643–663, Aug. 2006, doi: 10.1007/s00421-006-0238-1.

[12]	J. C. Strickland and M. A. Smith, “The anxiolytic effects of resistance exercise,” Front. Psychol., vol. 5, Jul. 2014, doi: 10.3389/fpsyg.2014.00753.

[13]	D. A. Kallman, C. C. Plato, and J. D. Tobin, “The Role of Muscle Loss in the Age-Related Decline of Grip Strength: Cross-Sectional and Longitudinal Perspectives,” J. Gerontol., vol. 45, no. 3, pp. M82–M88, May 1990, doi: 10.1093/geronj/45.3.M82.

[14]	M. D. Peterson and P. M. Gordon, “Resistance Exercise for the Aging Adult: Clinical Implications and Prescription Guidelines,” Am. J. Med., vol. 124, no. 3, pp. 194–198, Mar. 2011, doi: 10.1016/j.amjmed.2010.08.020.

[15]	S. Melov, M. A. Tarnopolsky, K. Beckman, K. Felkey, and A. Hubbard, “Resistance Exercise Reverses Aging in Human Skeletal Muscle,” PLoS ONE, vol. 2, no. 5, p. e465, May 2007, doi: 10.1371/journal.pone.0000465.

[16]	T. W. STORER, “Exercise in chronic pulmonary disease: resistance exercise prescription,” Med. Sci. Sports Exerc., vol. 33, no. Supplement, pp. S680–S686, Jul. 2001, doi: 10.1097/00005768-200107001-00006.

[17]	W. J. Evans, “Skeletal muscle loss: cachexia, sarcopenia, and inactivity,” Am. J. Clin. Nutr., vol. 91, no. 4, pp. 1123S-1127S, Apr. 2010, doi: 10.3945/ajcn.2010.28608A.

[18]	F. Landi et al., “Muscle loss: The new malnutrition challenge in clinical practice,” Clin. Nutr., vol. 38, no. 5, pp. 2113–2120, Oct. 2019, doi: 10.1016/j.clnu.2018.11.021.

[19]	M. DÁTTILO et al., “Effects of sleep deprivation on the Acute Skeletal Muscle Recovery after
Exercise,” Med. Sci. Sports Exerc., vol. 52, no. 2, pp. 507–514, Feb. 2020, doi: 10.1249/MSS.0000000000002137.

[20]	M. Dattilo et al., “Sleep and muscle recovery: Endocrinological and molecular basis for a new and promising hypothesis,” Med. Hypotheses, vol. 77, no. 2, pp. 220–222, Aug. 2011, doi: 10.1016/j.mehy.2011.04.017.

[21]	W. Lin et al., “The Effect of Sleep Restriction, With or Without Exercise, on Skeletal Muscle Transcriptomic Profiles in Healthy Young Males,” Front. Endocrinol., vol. 13, Jul. 2022, doi: 10.3389/fendo.2022.863224.

[22]	L. de S. N. Freitas et al., “Sleep debt induces skeletal muscle injuries in athletes: A promising hypothesis,” Med. Hypotheses, vol. 142, p. 109836, Sep. 2020, doi: 10.1016/j.mehy.2020.109836.

[23]	N. J. Saner et al., “The effect of sleep restriction, with or without
high-intensity interval exercise, on myofibrillar
protein synthesis in healthy young men,” J. Physiol., vol. 598, no. 8, pp. 1523–1536, Apr. 2020, doi: 10.1113/JP278828.

[24]	H. S. Driver and S. R. Taylor, “Exercise and sleep,” Sleep Med. Rev., vol. 4, no. 4, pp. 387–402, Aug. 2000, doi: 10.1053/smrv.2000.0110.

[25]	S. Uchida, K. Shioda, Y. Morita, C. Kubota, M. Ganeko, and N. Takeda, “Exercise Effects on Sleep Physiology,” Front. Neurol., vol. 3, 2012, doi: 10.3389/fneur.2012.00048.

[26]	J. Vierck et al., “SATELLITE CELL REGULATION FOLLOWING MYOTRAUMA CAUSED BY RESISTANCE EXERCISE,” Cell Biol. Int., vol. 24, no. 5, pp. 263–272, May 2000, doi: 10.1006/cbir.2000.0499.

[27]	S. Schiaffino, K. A. Dyar, S. Ciciliot, B. Blaauw, and M. Sandri, “Mechanisms regulating skeletal muscle growth and atrophy,” FEBS J., vol. 280, no. 17, pp. 4294–4314, Sep. 2013, doi: 10.1111/febs.12253.

[28]	B. C. Berk, “Vascular Smooth Muscle Growth: Autocrine Growth Mechanisms,” Physiol. Rev., vol. 81, no. 3, pp. 999–1030, Jul. 2001, doi: 10.1152/physrev.2001.81.3.999.

[29]	G. Pallafacchina, B. Blaauw, and S. Schiaffino, “Role of satellite cells in muscle growth and maintenance of muscle mass,” Nutr. Metab. Cardiovasc. Dis., vol. 23, pp. S12–S18, Dec. 2013, doi: 10.1016/j.numecd.2012.02.002.

[30]	J. G. Tidball, “Mechanical signal transduction in skeletal muscle growth and adaptation,” J. Appl. Physiol., vol. 98, no. 5, pp. 1900–1908, May 2005, doi: 10.1152/japplphysiol.01178.2004.

[31]	O. Keskin, A. Gursoy, B. Ma, and R. Nussinov, “Principles of Protein−Protein Interactions: What are the Preferred Ways For Proteins To Interact?,” Chem. Rev., vol. 108, no. 4, pp. 1225–1244, Apr. 2008, doi: 10.1021/cr040409x.

[32]	G. Wu, “Dietary protein intake and human health,” Food Funct., vol. 7, no. 3, pp. 1251–1265, 2016, doi: 10.1039/C5FO01530H.

[33]	M. S. Westerterp-Plantenga, A. Nieuwenhuizen, D. Tomé, S. Soenen, and K. R. Westerterp, “Dietary Protein, Weight Loss, and Weight Maintenance,” Annu. Rev. Nutr., vol. 29, no. 1, pp. 21–41, Aug. 2009, doi: 10.1146/annurev-nutr-080508-141056.

[34]	D. J. Millward, P. J. Garlick, R. J. C. Stewart, D. O. Nnanyelugo, and J. C. Waterlow, “Skeletal-muscle growth and protein turnover,” Biochem. J., vol. 150, no. 2, pp. 235–243, Aug. 1975, doi: 10.1042/bj1500235.

[35]	J.-P. Bonjour, “Dietary Protein: An Essential Nutrient For Bone Health,” J. Am. Coll. Nutr., vol. 24, no. sup6, pp. 526S-536S, Dec. 2005, doi: 10.1080/07315724.2005.10719501.

[36]	S. M. Phillips and L. J. C. V. Loon, “Dietary protein for athletes: From requirements to optimum adaptation,” J. Sports Sci., vol. 29, no. sup1, pp. S29–S38, Jan. 2011, doi: 10.1080/02640414.2011.619204.

[37]	E. T. Olaniyan, F. O’Halloran, and A. L. McCarthy, “Dietary protein considerations for muscle protein synthesis and muscle mass preservation in older adults,” Nutr. Res. Rev., vol. 34, no. 1, pp. 147–157, Jun. 2021, doi: 10.1017/S0954422420000219.

[38]	J. D. Kingsley and A. Figueroa, “Acute and training effects of resistance exercise on heart rate variability,” Clin. Physiol. Funct. Imaging, vol. 36, no. 3, pp. 179–187, May 2016, doi: 10.1111/cpf.12223.

[39]	J. M. Peake, O. Neubauer, P. A. D. Gatta, and K. Nosaka, “Muscle damage and inflammation during recovery from exercise,” J. Appl. Physiol., vol. 122, no. 3, pp. 559–570, Mar. 2017, doi: 10.1152/japplphysiol.00971.2016.

[40]	B. K. PEDERSEN, T. ROHDE, and K. OSTROWSKI, “Recovery of the immune system after exercise,” Acta Physiol. Scand., vol. 162, no. 3, pp. 325–332, Feb. 1998, doi: 10.1046/j.1365-201X.1998.0325e.x.

[41]	I. M. Wilcock, J. B. Cronin, and W. A. Hing, “Water Immersion: Does It Enhance Recovery From Exercise?,” Int. J. Sports Physiol. Perform., vol. 1, no. 3, pp. 195–206, Sep. 2006, doi: 10.1123/ijspp.1.3.195.

[42]	L. Smolak, S. K. Murnen, and J. K. Thompson, “Sociocultural Influences and Muscle Building in Adolescent Boys.,” Psychol. Men Masculinity, vol. 6, no. 4, pp. 227–239, Oct. 2005, doi: 10.1037/1524-9220.6.4.227.

[43]	F. Damas et al., “Resistance training‐induced changes in integrated myofibrillar protein synthesis are related to hypertrophy only after attenuation of muscle damage,” J. Physiol., vol. 594, no. 18, pp. 5209–5222, Sep. 2016, doi: 10.1113/JP272472.

[44]	J. D. MacDougall, M. J. Gibala, M. A. Tarnopolsky, J. R. MacDonald, S. A. Interisano, and K. E. Yarasheski, “The Time Course for Elevated Muscle Protein Synthesis Following Heavy Resistance Exercise,” Can. J. Appl. Physiol., vol. 20, no. 4, pp. 480–486, Dec. 1995, doi: 10.1139/h95-038.

[45]	H. C. Dreyer, S. Fujita, E. L. Glynn, M. J. Drummond, E. Volpi, and B. B. Rasmussen, “Resistance exercise increases leg muscle protein synthesis and mTOR signalling independent of sex,” Acta Physiol., vol. 199, no. 1, pp. 71–81, May 2010, doi: 10.1111/j.1748-1716.2010.02074.x.

[46]	V. Kumar et al., “Age‐related differences in the dose-response relationship of muscle protein synthesis to resistance exercise in young and old men,” J. Physiol., vol. 587, no. 1, pp. 211–217, Jan. 2009, doi: 10.1113/jphysiol.2008.164483.

[47]	S. M. Phillips, K. D. Tipton, A. A. Ferrando, and R. R. Wolfe, “Resistance training reduces the acute exercise-induced increase in muscle protein turnover,” Am. J. Physiol.-Endocrinol. Metab., vol. 276, no. 1, pp. E118–E124, Jan. 1999, doi: 10.1152/ajpendo.1999.276.1.E118.

[48]	J. N. Schulte and K. E. Yarasheski, “Effects of Resistance Training on the Rate of Muscle Protein Synthesis in Frail Elderly People,” Int. J. Sport Nutr. Exerc. Metab., vol. 11, no. s1, pp. S111–S118, Dec. 2001, doi: 10.1123/ijsnem.11.s1.s111.

[49]	K. E. Yarasheski, J. J. Zachwieja, and D. M. Bier, “Acute effects of resistance exercise on muscle protein synthesis rate in young and elderly men and women,” Am. J. Physiol.-Endocrinol. Metab., vol. 265, no. 2, pp. E210–E214, Aug. 1993, doi: 10.1152/ajpendo.1993.265.2.E210.

[50]	H. G. Gasier, S. E. Riechman, M. P. Wiggs, A. Buentello, S. F. Previs, and J. D. Fluckey, “Cumulative responses of muscle protein synthesis are augmented with chronic resistance exercise training,” Acta Physiol., vol. 201, no. 3, pp. 381–389, Mar. 2011, doi: 10.1111/j.1748-1716.2010.02183.x.

[51]	A. M. Holwerda et al., “Daily resistance-type exercise stimulates muscle protein synthesis in vivo in young men,” J. Appl. Physiol., vol. 124, no. 1, pp. 66–75, Jan. 2018, doi: 10.1152/japplphysiol.00610.2017.

[52]	J. E. Tang, J. G. Perco, D. R. Moore, S. B. Wilkinson, and S. M. Phillips, “Resistance training alters the response of fed state mixed muscle protein synthesis in young men,” Am. J. Physiol.-Regul. Integr. Comp. Physiol., vol. 294, no. 1, pp. R172–R178, Jan. 2008, doi: 10.1152/ajpregu.00636.2007.

[53]	A. A. Ferrando, K. D. Tipton, M. M. Bamman, and R. R. Wolfe, “Resistance exercise maintains skeletal muscle protein synthesis during bed rest,” J. Appl. Physiol., vol. 82, no. 3, pp. 807–810, Mar. 1997, doi: 10.1152/jappl.1997.82.3.807.

[54]	N. A. Burd et al., “Low-Load High Volume Resistance Exercise Stimulates Muscle Protein Synthesis More Than High-Load Low Volume Resistance Exercise in Young Men,” PLoS ONE, vol. 5, no. 8, p. e12033, Aug. 2010, doi: 10.1371/journal.pone.0012033.

[55]	S. Welle, C. Thornton, and M. Statt, “Myofibrillar protein synthesis in young and old human subjects after three months of resistance training,” Am. J. Physiol.-Endocrinol. Metab., vol. 268, no. 3, pp. E422–E427, Mar. 1995, doi: 10.1152/ajpendo.1995.268.3.E422.

[56]	S. M. Phillips, K. D. Tipton, A. Aarsland, S. E. Wolf, and R. R. Wolfe, “Mixed muscle protein synthesis and breakdown after resistance exercise in humans,” Am. J. Physiol.-Endocrinol. Metab., vol. 273, no. 1, pp. E99–E107, Jul. 1997, doi: 10.1152/ajpendo.1997.273.1.E99.

[57]	C. J. Mitchell et al., “Acute Post-Exercise Myofibrillar Protein Synthesis Is Not Correlated with Resistance Training-Induced Muscle Hypertrophy in Young Men,” PLoS ONE, vol. 9, no. 2, p. e89431, Feb. 2014, doi: 10.1371/journal.pone.0089431.

[58]	D. L. Mayhew, J. Kim, J. M. Cross, A. A. Ferrando, and M. M. Bamman, “Translational signaling responses preceding resistance training-mediated myofiber hypertrophy in young and old humans,” J. Appl. Physiol., vol. 107, no. 5, pp. 1655–1662, Nov. 2009, doi: 10.1152/japplphysiol.91234.2008.

[59]	O. C. WITARD, M. TIELAND, M. BEELEN, K. D. TIPTON, L. J. C. V. LOON, and R. KOOPMAN, “Resistance Exercise Increases Postprandial Muscle Protein Synthesis in Humans,” Med. Sci. Sports Exerc., vol. 41, no. 1, pp. 144–154, Jan. 2009, doi: 10.1249/MSS.0b013e3181844e79.

[60]	D. L. Hasten, J. Pak-Loduca, K. A. Obert, and K. E. Yarasheski, “Resistance exercise acutely increases MHC and mixed muscle protein synthesis rates in 78–84 and 23–32 yr olds,” Am. J. Physiol.-Endocrinol. Metab., vol. 278, no. 4, pp. E620–E626, Apr. 2000, doi: 10.1152/ajpendo.2000.278.4.E620.

[61]	N. A. Burd et al., “Resistance exercise volume affects myofibrillar protein synthesis and anabolic signalling molecule phosphorylation in young men,” J. Physiol., vol. 588, no. 16, pp. 3119–3130, Aug. 2010, doi: 10.1113/jphysiol.2010.192856.

[62]	H. C. Dreyer, S. Fujita, J. G. Cadenas, D. L. Chinkes, E. Volpi, and B. B. Rasmussen, “Resistance exercise increases AMPK activity and reduces 4E‐BP1 phosphorylation and protein synthesis in human skeletal muscle,” J. Physiol., vol. 576, no. 2, pp. 613–624, Oct. 2006, doi: 10.1113/jphysiol.2006.113175.

[63]	N. A. Burd et al., “Muscle time under tension during resistance exercise stimulates differential muscle protein sub‐fractional synthetic responses in men,” J. Physiol., vol. 590, no. 2, pp. 351–362, Jan. 2012, doi: 10.1113/jphysiol.2011.221200.

[64]	S. B. Wilkinson et al., “Differential effects of resistance and endurance exercise in the fed state on signalling molecule phosphorylation and protein synthesis in human muscle,” J. Physiol., vol. 586, no. 15, pp. 3701–3717, Aug. 2008, doi: 10.1113/jphysiol.2008.153916.

[65]	P. L. Kim, R. S. Staron, and S. M. Phillips, “Fasted‐state skeletal muscle protein synthesis after resistance exercise is altered with training,” J. Physiol., vol. 568, no. 1, pp. 283–290, Oct. 2005, doi: 10.1113/jphysiol.2005.093708.

[66]	A. Chesley, J. D. MacDougall, M. A. Tarnopolsky, S. A. Atkinson, and K. Smith, “Changes in human muscle protein synthesis after resistance exercise,” J. Appl. Physiol., vol. 73, no. 4, pp. 1383–1388, Oct. 1992, doi: 10.1152/jappl.1992.73.4.1383.

[67]	K. E. Yarasheski, J. Pak-Loduca, D. L. Hasten, K. A. Obert, M. B. Brown, and D. R. Sinacore, “Resistance exercise training increases mixed muscle protein synthesis rate in frail women and men ≥76 yr old,” Am. J. Physiol.-Endocrinol. Metab., vol. 277, no. 1, pp. E118–E125, Jul. 1999, doi: 10.1152/ajpendo.1999.277.1.E118.

[68]	S. Fujita et al., “Blood flow restriction during low-intensity resistance exercise increases S6K1 phosphorylation and muscle protein synthesis,” J. Appl. Physiol., vol. 103, no. 3, pp. 903–910, Sep. 2007, doi: 10.1152/japplphysiol.00195.2007.

[69]	K. E. Yarasheski, J. J. Zachwieja, J. A. Campbell, and D. M. Bier, “Effect of growth hormone and resistance exercise on muscle growth and strength in older men,” Am. J. Physiol.-Endocrinol. Metab., vol. 268, no. 2, pp. E268–E276, Feb. 1995, doi: 10.1152/ajpendo.1995.268.2.E268.

[70]	N. A. Burd et al., “Enhanced Amino Acid Sensitivity of Myofibrillar Protein Synthesis Persists for up to 24 h after Resistance Exercise in Young Men1–3,” J. Nutr., vol. 141, no. 4, pp. 568–573, Apr. 2011, doi: 10.3945/jn.110.135038.

[71]	C. S. Fry et al., “Skeletal Muscle Autophagy and Protein Breakdown Following Resistance Exercise are Similar in Younger and Older Adults,” J. Gerontol. A. Biol. Sci. Med. Sci., vol. 68, no. 5, pp. 599–607, May 2013, doi: 10.1093/gerona/gls209.

[72]	K. E. Bell, C. Séguin, G. Parise, S. K. Baker, and S. M. Phillips, “Day-to-Day Changes in Muscle Protein Synthesis in Recovery From Resistance, Aerobic, and High-Intensity Interval Exercise in Older Men,” J. Gerontol. A. Biol. Sci. Med. Sci., vol. 70, no. 8, pp. 1024–1029, Aug. 2015, doi: 10.1093/gerona/glu313.

[73]	C. J. Hulston et al., “Resistance exercise stimulates mixed muscle protein synthesis in lean and obese young adults,” Physiol. Rep., vol. 6, no. 14, p. e13799, Jul. 2018, doi: 10.14814/phy2.13799.

[74]	M. J. Drummond et al., “Skeletal muscle protein anabolic response to resistance exercise and essential amino acids is delayed with aging,” J. Appl. Physiol., vol. 104, no. 5, pp. 1452–1461, May 2008, doi: 10.1152/japplphysiol.00021.2008.

[75]	C. J. Mitchell, T. A. Churchward-Venne, L. Bellamy, G. Parise, S. K. Baker, and S. M. Phillips, “Muscular and Systemic Correlates of Resistance Training-Induced Muscle Hypertrophy,” PLoS ONE, vol. 8, no. 10, p. e78636, Oct. 2013, doi: 10.1371/journal.pone.0078636.

[76]	J. McKendry et al., “Short inter‐set rest blunts resistance exercise‐induced increases in myofibrillar protein synthesis and intracellular signalling in young males,” Exp. Physiol., vol. 101, no. 7, pp. 866–882, Jul. 2016, doi: 10.1113/EP085647.

[77]	L. S. Macnaughton et al., “The response of muscle protein synthesis following whole‐body resistance exercise is greater following 40 g than 20 g of ingested whey protein,” Physiol. Rep., vol. 4, no. 15, Aug. 2016, doi: 10.14814/phy2.12893.

[78]	Oliver’ ’Witard, Sarah’ ’Jackman, Leigh’ ’Breen, Kenneth’ ’Smith, Anna’ ’Selby, and Kevin’ ’Tipton, “Myofibrillar muscle protein synthesis rates subsequent to a meal in response to increasing doses of whey protein at rest and after resistance exercise,” no. 2013, Oct. 2012.

[79]	D. R. Moore et al., “Ingested protein dose response of muscle and albumin protein synthesis after resistance exercise in young men,” Am. J. Clin. Nutr., vol. 89, no. 1, pp. 161–168, Jan. 2009, doi: 10.3945/ajcn.2008.26401.

[80]	B. J. Schoenfeld and A. A. Aragon, “How much protein can the body use in a single meal for muscle-building? Implications for daily protein distribution,” J. Int. Soc. Sports Nutr., vol. 15, no. 1, Jan. 2018, doi: 10.1186/s12970-018-0215-1.

[81]	J. L. Areta et al., “Reduced resting skeletal muscle protein synthesis is rescued by resistance exercise and protein ingestion following short-term energy deficit,” Am. J. Physiol.-Endocrinol. Metab., vol. 306, no. 8, pp. E989–E997, Apr. 2014, doi: 10.1152/ajpendo.00590.2013.

[82]	J. L. Areta et al., “Timing and distribution of protein ingestion during prolonged recovery from resistance exercise alters myofibrillar protein synthesis,” J. Physiol., vol. 591, no. 9, pp. 2319–2331, May 2013, doi: 10.1113/jphysiol.2012.244897.

[83]	D. W. West et al., “Rapid aminoacidemia enhances myofibrillar protein synthesis and anabolic intramuscular signaling responses after resistance exercise,” Am. J. Clin. Nutr., vol. 94, no. 3, pp. 795–803, Sep. 2011, doi: 10.3945/ajcn.111.013722.

[84]	B. Esmarck, J. L. Andersen, S. Olsen, E. A. Richter, M. Mizuno, and M. Kjær, “Timing of postexercise protein intake is important for muscle hypertrophy with resistance training in elderly humans,” J. Physiol., vol. 535, no. 1, pp. 301–311, Aug. 2001, doi: 10.1111/j.1469-7793.2001.00301.x.

[85]	T. A. Churchward-Venne et al., “Dose-response effects of dietary protein on muscle protein synthesis during recovery from endurance exercise in young men: a double-blind randomized trial,” Am. J. Clin. Nutr., vol. 112, no. 2, pp. 303–317, Aug. 2020, doi: 10.1093/ajcn/nqaa073.

[86]	K. J. Dideriksen et al., “Stimulation of muscle protein synthesis by whey and caseinate ingestion after resistance exercise in elderly individuals,” Scand. J. Med. Sci. Sports, vol. 21, no. 6, Dec. 2011, doi: 10.1111/j.1600-0838.2011.01318.x.

[87]	K. D. Tipton, T. A. Elliott, M. G. Cree, A. A. Aarsland, A. P. Sanford, and R. R. Wolfe, “Stimulation of net muscle protein synthesis by whey protein ingestion before and after exercise,” Am. J. Physiol.-Endocrinol. Metab., vol. 292, no. 1, pp. E71–E76, Jan. 2007, doi: 10.1152/ajpendo.00166.2006.

[88]	T. A. Churchward‐Venne et al., “Supplementation of a suboptimal protein dose with leucine or essential amino acids: effects on myofibrillar protein synthesis at rest and following resistance exercise in men,” J. Physiol., vol. 590, no. 11, pp. 2751–2765, Jun. 2012, doi: 10.1113/jphysiol.2012.228833.

[89]	K. L. English and D. Paddon-Jones, “Protecting muscle mass and function in older adults during bed rest,” Curr. Opin. Clin. Nutr. Metab. Care, vol. 13, no. 1, pp. 34–39, Jan. 2010, doi: 10.1097/MCO.0b013e328333aa66.

[90]	Y. Yang, T. A. Churchward-Venne, N. A. Burd, L. Breen, M. A. Tarnopolsky, and S. M. Phillips, “Myofibrillar protein synthesis following ingestion of soy protein isolate at rest and after resistance exercise in elderly men,” Nutr. Metab., vol. 9, no. 1, p. 57, Dec. 2012, doi: 10.1186/1743-7075-9-57.

[91]	D. S. Willoughby, J. R. Stout, and C. D. Wilborn, “Effects of resistance training and protein plus amino acid supplementation on muscle anabolism, mass, and strength,” Amino Acids, vol. 32, no. 4, pp. 467–477, May 2007, doi: 10.1007/s00726-006-0398-7.

[92]	D. R. Moore, J. E. Tang, N. A. Burd, T. Rerecich, M. A. Tarnopolsky, and S. M. Phillips, “Differential stimulation of myofibrillar and sarcoplasmic protein synthesis with protein ingestion at rest and after resistance exercise,” J. Physiol., vol. 587, no. 4, pp. 897–904, Feb. 2009, doi: 10.1113/jphysiol.2008.164087.

[93]	S. Y. Oikawa et al., “A randomized controlled trial of the impact of protein supplementation on leg lean mass and integrated muscle protein synthesis during inactivity and energy restriction in older persons,” Am. J. Clin. Nutr., vol. 108, no. 5, pp. 1060–1068, Nov. 2018, doi: 10.1093/ajcn/nqy193.

[94]	L. L. Andersen et al., “The effect of resistance training combined with timed ingestion of protein on muscle fiber size and muscle strength,” Metabolism, vol. 54, no. 2, pp. 151–156, Feb. 2005, doi: 10.1016/j.metabol.2004.07.012.

[95]	E. Børsheim, K. D. Tipton, S. E. Wolf, and R. R. Wolfe, “Essential amino acids and muscle protein recovery from resistance exercise,” Am. J. Physiol.-Endocrinol. Metab., vol. 283, no. 4, pp. E648–E657, Oct. 2002, doi: 10.1152/ajpendo.00466.2001.

[96]	J. E. Tang, J. J. Manolakos, G. W. Kujbida, P. J. Lysecki, D. R. Moore, and S. M. Phillips, “Minimal whey protein with carbohydrate stimulates muscle protein synthesis following resistance exercise in trained young men,” Appl. Physiol. Nutr. Metab., vol. 32, no. 6, pp. 1132–1138, Dec. 2007, doi: 10.1139/H07-076.

[97]	K. R. Howarth, N. A. Moreau, S. M. Phillips, and M. J. Gibala, “Coingestion of protein with carbohydrate during recovery from endurance exercise stimulates skeletal muscle protein synthesis in humans,” J. Appl. Physiol., vol. 106, no. 4, pp. 1394–1402, Apr. 2009, doi: 10.1152/japplphysiol.90333.2008.

[98]	J. E. Tang, D. R. Moore, G. W. Kujbida, M. A. Tarnopolsky, and S. M. Phillips, “Ingestion of whey hydrolysate, casein, or soy protein isolate: effects on mixed muscle protein synthesis at rest and following resistance exercise in young men,” J. Appl. Physiol., vol. 107, no. 3, pp. 987–992, Sep. 2009, doi: 10.1152/japplphysiol.00076.2009.

[99]	S. Fujita, H. C. Dreyer, M. J. Drummond, E. L. Glynn, E. Volpi, and B. B. Rasmussen, “Essential amino acid and carbohydrate ingestion before resistance exercise does not enhance postexercise muscle protein synthesis,” J. Appl. Physiol., vol. 106, no. 5, pp. 1730–1739, May 2009, doi: 10.1152/japplphysiol.90395.2008.

[100]	S. M. Pasiakos et al., “Effects of high‐protein diets on fat‐free mass and muscle protein synthesis following weight loss: a randomized controlled trial,” FASEB J., vol. 27, no. 9, pp. 3837–3847, Sep. 2013, doi: 10.1096/fj.13-230227.

[101]	C. Roth, L. Rettenmaier, and M. Behringer, “High-Protein Energy-Restriction: Effects on Body Composition, Contractile Properties, Mood, and Sleep in Active Young College Students,” Front. Sports Act. Living, vol. 3, Jun. 2021, doi: 10.3389/fspor.2021.683327.

[102]	R. Koopman et al., “Co-ingestion of protein and leucine stimulates muscle protein synthesis rates to the same extent in young and elderly lean men,” Am. J. Clin. Nutr., vol. 84, no. 3, pp. 623–632, Dec. 2006, doi: 10.1093/ajcn/84.3.623.

[103]	B. Pennings, R. Koopman, M. Beelen, J. M. Senden, W. H. Saris, and L. J. van Loon, “Exercising before protein intake allows for greater use of dietary protein–derived amino acids for de novo muscle protein synthesis in both young and elderly men,” Am. J. Clin. Nutr., vol. 93, no. 2, pp. 322–331, Feb. 2011, doi: 10.3945/ajcn.2010.29649.

[104]	C. Mitchell et al., “Consumption of Milk Protein or Whey Protein Results in a Similar Increase in Muscle Protein Synthesis in Middle Aged Men,” Nutrients, vol. 7, no. 10, pp. 8685–8699, Oct. 2015, doi: 10.3390/nu7105420.

[105]	S. K. Rahbek et al., “Effects of divergent resistance exercise contraction mode and dietary supplementation type on anabolic signalling, muscle protein synthesis and muscle hypertrophy,” Amino Acids, vol. 46, no. 10, pp. 2377–2392, Oct. 2014, doi: 10.1007/s00726-014-1792-1.

[106]	M. Beelen et al., “Coingestion of Carbohydrate and Protein Hydrolysate Stimulates Muscle Protein Synthesis during Exercise in Young Men, with No Further Increase during Subsequent Overnight Recovery,” J. Nutr., vol. 138, no. 11, pp. 2198–2204, Nov. 2008, doi: 10.3945/jn.108.092924.

[107]	J. Bülow et al., “Effect of 1-year daily protein supplementation and physical exercise on muscle protein synthesis rate and muscle metabolome in healthy older Danes: a randomized controlled trial,” Eur. J. Nutr., vol. 62, no. 6, pp. 2673–2685, Sep. 2023, doi: 10.1007/s00394-023-03182-0.

[108]	T. B. Symons, M. Sheffield-Moore, R. R. Wolfe, and D. Paddon-Jones, “A Moderate Serving of High-Quality Protein Maximally Stimulates Skeletal Muscle Protein Synthesis in Young and Elderly Subjects,” J. Am. Diet. Assoc., vol. 109, no. 9, pp. 1582–1586, Sep. 2009, doi: 10.1016/j.jada.2009.06.369.

[109]	G. I. Smith, P. K. Commean, D. N. Reeds, S. Klein, and B. Mittendorfer, “Effect of Protein Supplementation During Diet‐Induced Weight Loss on Muscle Mass and Strength: A Randomized Controlled Study,” Obesity, vol. 26, no. 5, pp. 854–861, May 2018, doi: 10.1002/oby.22169.

[110]	Y. C. Luiking, N. E. Deutz, R. G. Memelink, S. Verlaan, and R. R. Wolfe, “Postprandial muscle protein synthesis is higher after a high whey protein, leucine-enriched supplement than after a dairy-like product in healthy older people: a randomized controlled trial,” Nutr. J., vol. 13, no. 1, p. 9, Dec. 2014, doi: 10.1186/1475-2891-13-9.

[111]	A. Kanda et al., “Post-exercise whey protein hydrolysate supplementation induces a greater increase in muscle protein synthesis than its constituent amino acid content,” Br. J. Nutr., vol. 110, no. 6, pp. 981–987, Sep. 2013, doi: 10.1017/S0007114512006174.

[112]	M. J. Ormsbee, P. G. Saracino, M. C. Morrissey, J. Donaldson, L. I. Rentería, and A. J. McKune, “Pre-sleep protein supplementation after an acute bout of evening resistance exercise does not improve next day performance or recovery in resistance trained men,” J. Int. Soc. Sports Nutr., vol. 19, no. 1, pp. 164–178, Dec. 2022, doi: 10.1080/15502783.2022.2036451.

[113]	C. E. G. Reis, L. M. R. Loureiro, H. Roschel, and T. H. M. da Costa, “Effects of pre-sleep protein consumption on muscle-related outcomes — A systematic review,” J. Sci. Med. Sport, vol. 24, no. 2, pp. 177–182, Feb. 2021, doi: 10.1016/j.jsams.2020.07.016.

[114]	D. Ramos-Campo, L. M. Martínez-Aranda, L. A. Caravaca, V. Ávila-Gandí, and J. Á. Rubio-Arias, “Effects of resistance training intensity on the sleep quality and strength recovery in trained men: a randomized cross-over study,” Biol. Sport, vol. 38, no. 1, pp. 81–88, 2021, doi: 10.5114/biolsport.2020.97677.

[115]	T. Snijders et al., “Protein Ingestion before Sleep Increases Muscle Mass and Strength Gains during Prolonged Resistance-Type Exercise Training in Healthy Young MenNitrogen1–3,” J. Nutr., vol. 145, no. 6, pp. 1178–1184, Jun. 2015, doi: 10.3945/jn.114.208371.

[116]	M. Morrison, S. L. Halson, J. Weakley, and J. A. Hawley, “Sleep, circadian biology and skeletal muscle interactions: Implications for metabolic health,” Sleep Med. Rev., vol. 66, p. 101700, Dec. 2022, doi: 10.1016/j.smrv.2022.101700.

[117]	J. A. Betts, M. Beelen, K. A. Stokes, W. H. M. Saris, and L. J. C. van Loon, “Endocrine Responses During Overnight Recovery From Exercise: Impact of Nutrition and Relationships With Muscle Protein Synthesis,” Int. J. Sport Nutr. Exerc. Metab., vol. 21, no. 5, pp. 398–409, Oct. 2011, doi: 10.1123/ijsnem.21.5.398.

[118]	P. G. Saracino, H. E. Saylor, B. R. Hanna, R. C. Hickner, J.-S. Kim, and M. J. Ormsbee, “Effects of Pre-Sleep Whey vs. Plant-Based Protein Consumption on Muscle Recovery Following Damaging Morning Exercise,” Nutrients, vol. 12, no. 7, p. 2049, Jul. 2020, doi: 10.3390/nu12072049.

[119]	I. W. Kouw et al., “Protein Ingestion before Sleep Increases Overnight Muscle Protein Synthesis Rates in Healthy Older Men: A Randomized Controlled Trial,” J. Nutr., vol. 147, no. 12, pp. 2252–2261, Dec. 2017, doi: 10.3945/jn.117.254532.

[120]	S. Lamon et al., “The effect of acute sleep deprivation on skeletal muscle protein synthesis and the hormonal environment,” 
Physiol. Rep., vol. 9, no. 1, Jan. 2021, doi: 10.14814/phy2.14660.

[121]	N. J. Saner et al., “Exercise mitigates sleep-loss-induced changes in glucose tolerance, mitochondrial function, sarcoplasmic protein synthesis, and diurnal rhythms,” Mol. Metab., vol. 43, p. 101110, Jan. 2021, doi: 10.1016/j.molmet.2020.101110.

[122]	A. M. Holwerda et al., “Physical Activity Performed in the Evening Increases the Overnight Muscle Protein Synthetic Response to Presleep Protein Ingestion in Older Men,” J. Nutr., vol. 146, no. 7, pp. 1307–1314, Jul. 2016, doi: 10.3945/jn.116.230086.

[123]	P. T. RES et al., “Protein Ingestion before Sleep Improves Postexercise Overnight Recovery,” Med. Sci. Sports Exerc., vol. 44, no. 8, pp. 1560–1569, Aug. 2012, doi: 10.1249/MSS.0b013e31824cc363.

[124]	B. B. L. Groen et al., “Intragastric protein administration stimulates overnight muscle protein synthesis in elderly men,” Am. J. Physiol.-Endocrinol. Metab., vol. 302, no. 1, pp. E52–E60, Jan. 2012, doi: 10.1152/ajpendo.00321.2011.

[125]	A. M. Holwerda et al., “ Protein Supplementation after Exercise and before Sleep Does Not Further Augment Muscle Mass and Strength Gains during Resistance Exercise Training in Active Older Men,” J. Nutr., vol. 148, no. 11, pp. 1723–1732, Nov. 2018, doi: 10.1093/jn/nxy169.

[126]	B. Smeuninx et al., “The effect of short‐term exercise prehabilitation on skeletal muscle protein synthesis and atrophy during bed rest in older men,” J. Cachexia Sarcopenia Muscle, vol. 12, no. 1, pp. 52–69, Feb. 2021, doi: 10.1002/jcsm.12661.

[127]	M. J. Drummond et al., “Bed rest impairs skeletal muscle amino acid transporter expression, mTORC1 signaling, and protein synthesis in response to essential amino acids in older adults,” Am. J. Physiol.-Endocrinol. Metab., vol. 302, no. 9, pp. E1113–E1122, May 2012, doi: 10.1152/ajpendo.00603.2011.

[128]	M. Mônico-Neto et al., “Resistance exercise: A non-pharmacological strategy to minimize or reverse sleep deprivation-induced muscle atrophy,” Med. Hypotheses, vol. 80, no. 6, pp. 701–705, Jun. 2013, doi: 10.1016/j.mehy.2013.02.013.

[129]	A. A. Ferrando, H. W. Lane, C. A. Stuart, J. Davis-Street, and R. R. Wolfe, “Prolonged bed rest decreases skeletal muscle and whole body protein synthesis,” Am. J. Physiol.-Endocrinol. Metab., vol. 270, no. 4, pp. E627–E633, Apr. 1996, doi: 10.1152/ajpendo.1996.270.4.E627.

[130]	L. C. Lyons, Y. Vanrobaeys, and T. Abel, “Sleep and memory: The impact of sleep deprivation on transcription, translational control, and protein synthesis in the brain,” J. Neurochem., vol. 166, no. 1, pp. 24–46, Jul. 2023, doi: 10.1111/jnc.15787.

[131]	F. M. Stich, S. Huwiler, G. D’Hulst, and C. Lustenberger, “The Potential Role of Sleep in Promoting a Healthy Body Composition: Underlying Mechanisms Determining Muscle, Fat, and Bone Mass and Their Association with Sleep,” Neuroendocrinology, vol. 112, no. 7, pp. 673–701, 2022, doi: 10.1159/000518691.

[132]	M. Chennaoui et al., “How does sleep help recovery from exercise-induced muscle injuries?” J. Sci. Med. Sport, vol. 24, no. 10, pp. 982–987, Oct. 2021, doi: 10.1016/j.jsams.2021.05.007.

[133]	N. Buchmann, D. Spira, K. Norman, I. Demuth, R. Eckardt, and E. Steinhagen-Thiessen, “Sleep, Muscle Mass and Muscle Function in Older People: A Cross-Sectional Analysis Based on Data From the Berlin Aging Study II (BASE-II),” Dtsch. Ärztebl. Int., Apr. 2016, doi: 10.3238/arztebl.2016.0253.

[134]	P. Kortebein, A. Ferrando, J. Lombeida, R. Wolfe, and W. J. Evans, “Effect of 10 Days of Bed Rest on Skeletal Muscle in Healthy Older Adults,” JAMA, vol. 297, no. 16, p. 1769, Apr. 2007, doi: 10.1001/jama.297.16.1772-b.

[135]	J. M. Peake, “Recovery after exercise: what is the current state of play?,” Curr. Opin. Physiol., vol. 10, pp. 17–26, Aug. 2019, doi: 10.1016/j.cophys.2019.03.007.

[136]	P. T. Reidy et al., “Neuromuscular Electrical Stimulation Combined with Protein Ingestion Preserves Thigh Muscle Mass But Not Muscle Function in Healthy Older Adults During 5 Days of Bed Rest,” Rejuvenation Res., vol. 20, no. 6, pp. 449–461, Dec. 2017, doi: 10.1089/rej.2017.1942.

[137]	V. A. R. Viana et al., “The effects of a session of resistance training on sleep patterns in the elderly,” Eur. J. Appl. Physiol., vol. 112, no. 7, pp. 2403–2408, Jul. 2012, doi: 10.1007/s00421-011-2219-2.

[138]	J. R. Alley, J. W. Mazzochi, C. J. Smith, D. M. Morris, and S. R. Collier, “Effects of Resistance Exercise Timing on Sleep Architecture and Nocturnal Blood Pressure,” J. Strength Cond. Res., vol. 29, no. 5, pp. 1378–1385, May 2015, doi: 10.1519/JSC.0000000000000750.

[139]	K. Shioda, K. Goto, and S. Uchida, “The effect of 2 consecutive days of intense resistance exercise on sleep in untrained adults,” Sleep Biol. Rhythms, vol. 17, no. 1, pp. 27–35, Jan. 2019, doi: 10.1007/s41105-018-0180-8.

[140]	D. J. Miller, C. Sargent, G. D. Roach, A. T. Scanlan, G. E. Vincent, and M. Lastella, “Moderate‐intensity exercise performed in the evening does not impair sleep in healthy males,” Eur. J. Sport Sci., vol. 20, no. 1, pp. 80–89, Feb. 2020, doi: 10.1080/17461391.2019.1611934.

[141]	P.-Y. Yang, K.-H. Ho, H.-C. Chen, and M.-Y. Chien, “Exercise training improves sleep quality in middle-aged and older adults with sleep problems: a systematic review,” J. Physiother., vol. 58, no. 3, pp. 157–163, Sep. 2012, doi: 10.1016/S1836-9553(12)70106-6.

[142]	M. Chennaoui, P. J. Arnal, F. Sauvet, and D. Léger, “Sleep and exercise: A reciprocal issue?” Sleep Med. Rev., vol. 20, pp. 59–72, Apr. 2015, doi: 10.1016/j.smrv.2014.06.008.

[143]	T. Madzima, J. Black, J. Melanson, S. Nepocatych, and E. Hall, “Influence of Resistance Exercise on Appetite and Affect Following Pre-Sleep Feeding,” Sports, vol. 6, no. 4, p. 172, Dec. 2018, doi: 10.3390/sports6040172.

\end{document}